\documentclass[12pt,a4paper]{article}
\usepackage{fullpage}
\usepackage{amsthm,latexsym,amsmath,amssymb,amsfonts,euscript,dsfont,mathrsfs}
\usepackage{mathtools}

\usepackage{authblk}
\usepackage{comment}

\usepackage{color}

\usepackage[section]{placeins}

\usepackage[latin1]{inputenc}
\usepackage{graphicx}
\usepackage[numbers]{natbib}

\def\P{\mathds{P}}
\def\E{\mathds{E}}
\def\V{\mathrm{Var}}
\def\A{\EuScript{A}}

\newcommand{\lawto}{\mathop{}\!\xrightarrow[t\to+\infty]{\mathrm{\scriptscriptstyle{law}}}}

\newcommand{\approxd}{\mathop{}\!\stackrel{\mathrm{\scriptscriptstyle{law}}}{\approx}}

 \newcommand{\dd}{\mathop{}\!\mathrm{d}}

\begin{document}

\title{Rejuvenating Functional Responses with Renewal Theory}

\author[1]{Sylvain Billiard
\thanks{ \scriptsize Email: \texttt{sylvain.billiard@univ-lille1.fr}}}
\affil[1]{\small Univ. Lille, CNRS, UMR 8198 - Evo-Eco-Paleo, F-59000 Lille, France}

\author[2]{Vincent Bansaye
\thanks{\scriptsize  Email: \texttt{bansaye@cmap.polytechnique.fr}}}
\affil[2]{Centre de Math\'ematiques Appliqu\'ees, CNRS UMR 7641, Ecole Polytechnique, F-91128 Palaiseau Cedex, France}

\author[3]{J.-R. Chazottes, 
\thanks{\scriptsize  Email: \texttt{chazottes@cpht.polytechnique.fr}}}
\affil[3]{Centre de Physique Th\'eorique, CNRS UMR 7644, Ecole Polytechnique, F-91128 Palaiseau Cedex, France}

\date{}

\maketitle 





{\par \bf Running title.} Stochastic derivation of functional responses



{\par \bf Corresponding author:} Sylvain Billiard: sylvain.billiard@univ-lille1.fr, Univ. Lille, CNRS, UMR 8198 - Evo-Eco-Paleo, F-59000 Lille, France.

\newpage

\begin{abstract}
Functional responses are widely used to describe interactions and resources exchange between individuals in ecology. The form given to functional responses dramatically affects the dynamics and stability of populations and communities. Despite their importance, functional responses are generally considered with a phenomenological approach, without clear mechanistic justifications from individual traits and behaviors. Here, we develop a bottom-up stochastic framework grounded in Renewal Theory showing how functional responses emerge from the level of the individuals through the decomposition of interactions into different activities. Our framework has many applications for conceptual, theoretical and empirical purposes. First, we show how the mean and variance of classical functional responses are obtained with explicit ecological assumptions, for instance regarding foraging behaviors. Second, we give examples in specific ecological contexts, such as in nuptial-feeding species or size dependent handling times. Finally, we demonstrate how to analyze data with our framework, especially highlighting that observed variability in the number of interactions can be used to infer parameters and compare functional response models.
\end{abstract}



\newpage

\section{Introduction}

Interactions between individuals affect all ecological processes, and how fast they occur determine resources exchange rates in an ecosystem. Interactions are generally considered at the population (or macroscopic) level and supposed to vary along with species densities following an interaction function, generally called a \emph{functional response}. The seminal papers introducing functional responses in population ecology followed a reductionistic approach and aimed at giving the underlying mechanisms \citep{holling1959, holling1965, denny2014}. Yet, population ecologists generally follow a phenomenological approach to justify functional responses \cite{jeschkeetal2002}, and functional responses are rarely derived from the individuals' traits and behaviors. The form given to functional responses is crucial since it can dramatically affect the dynamics and stability of populations and communities. For instance, the stability of predator and prey populations strongly depends on whether predator consumption rates increases linearly (Holling type I functional response) or following a saturating function (Holling Type II and III functional responses) with prey densities \citep{hastings1997}. It is thus critical to modelize within and between species interactions as accurately as possible in order to have the best predictions and understanding of population and community dynamics, and eventually support wildlife management decisions \citep{pettorellietal2015}. 

The form given to functional responses is largely debated for decades, sometimes fiercely (\emph{e.g.} the long-standing controversy about whether it is best to assume density-dependence or ratio-dependence in predator-prey models \citep{arditiginzburg2012, abrams2015}). Hundreds of functional responses have been proposed in the literature regarding all types of interactions: cooperation \citep{hollandetal2002}, plant-pollinators \citep{hollanddeangelis2010, fishmanhadany2010}, predation (reviewed in \cite{jeschkeetal2002, abrams2015}), competition \citep{johanssonsumpter2003}. Strikingly, despite a large variety of possible functional responses, Holling type II or related functional responses are most often used  \citep{jeschkeetal2002,baskett2012}, either for predator-prey \citep{skalskigilliam2001} or mutualistic interactions \citep{hollanddeangelis2010}. There is however a general agreement that there is much room for improvement. It is for instance difficult to determine which one of alternative functional response fits the best to empirical data because of the poor statistical power of fitting different functions to data \citep{skalskigilliam2001,gonzalezsuarezetal2011}. Holling Type II functional response is certainly preferred not because it adequately modelizes ecological interactions but rather because it is a saturating function with a single parameter. Many authors argue that bridging the gap between interactions at the level of the individuals (the microscopic scale) and functional responses (the macroscopic level) would be a critical milestone in the field \citep{pettorellietal2015, kalinoskianddelong2016}, first because it is necessary to mechanistically justify functional responses, and second because evidence of individual traits variation in functional responses accumulate \citep[\emph{e.g.}][]{pettorellietal2015, schroderetal2016}. 

Few studies aimed at making the link between microscopic and macroscopic scales in the context of functional responses. Some authors used a mean-field approximation to derive functional response in consumer-resources relationships  \cite[\emph{e.g.}][]{holling1959,pawaretal2012}. In particular, inspired by his famous ``Disc experiment'', Holling  showed that considering mechanisms such as searching times proportional to prey density and constant handling times of prey by predators, gives the well-known Holling's Type II functional response \cite{holling1959}. Holling later showed that Holling Type III functional response can be derived by introducing the mechanism of predators learning \cite{holling1965}. Other authors used deterministic approaches derived from chemical reactions equations to show which assumptions must be verified for a specific functional response to be valid, \textit{e.g.} Holling functional responses \citep{metzdiekmann1986}, Beddington-DeAngelis functional response \citep{geritzgyllenberg2012}, plant-marking pollinators interactions \citep{fishmanhadany2010}. However, these approaches have strong limitations. First, they can be used only to derive simple functional responses. Indeed, approximations of deterministic functional responses can be obtained from a system of ordinary differential equations under assumptions of slow/fast processes. Approximations might generally be difficult to obtain because the number of equations and simplifying assumptions to make increase with the complexity of the system. It makes this approach unlikely to be general enough to embrace the large variety of possible ecological contexts. Second, since only the mean rates of interaction at the population level are considered, variability of traits and behaviors between individuals can not be considered. Third, since these approaches are deterministic, they are only valid in very large populations and the importance of stochasticity can not be taken into account. 

Deriving stochastic models of functional responses have two purposes. First, it makes explicit the assumptions underlying the average interaction rates, \emph{i.e.} the mean of functional responses. It has been performed in few studies in specific situations. \cite{dawessouza2013} analyzed a Markov-chain model of a predator-prey system adapted from chemical reactions, and showed how Holling functional responses could emerge at the population level. \cite{broometal2010} and \cite{smallegangevandermeer2010} used a similar approach to derive a functional response in the specific cases of kleptoparasitism and a Beddington-DeAngelis functional response, respectively. Second, the development of stochastic models are needed for inferring functional responses from empirical data in order to clearly identify the processes and mechanisms underlying the variability of interaction rates, which appears to be large in experiments \cite[\emph{e.g.}][]{schroderetal2016}. Three sources of variability are possible: i) environmental or exogenous variability, \emph{e.g.} temperature or observation errors; ii) inter-individual variation of behavior or traits, \emph{e.g.} size, color or run speed; iii)  endogenous variability due to stochastic fluctuations of the interaction themselves, hereafter called \emph{interaction stochasticity}. 

 Modelling interactions processes at the microscopic level is necessary to evaluate the part of variance due to interaction stochasticity. \cite{johanssonsumpter2003} is the only study to our knowledge which proposed an explicit expression for an approximation of variance due to interactions, in the specific case of competition for resources. They assumed a site-based models where individuals are randomly distributed in patches every generation. They showed how different functional responses for competition can be derived depending on the assumptions about how resources are shared between competitors. A general stochastic framework for the derivation of functional responses is however still lacking. This would help to better justify the choice of the form given to functional responses, in interpreting data and making statistical inference, and evaluating the importance of interaction stochasticity for the dynamics and stability of populations and communities. In addition, since the variance due to interaction stochasticity comes from the interactions process itself, it is a valuable source of informations for inference, and would help for parameters estimation and models selection.      

In this paper, we propose a general stochastic framework which allows accounting for interaction mechanisms at the level of individuals and the derivation of functional responses at the population or community level. It is based on the modelling of the distribution of the times separating two interactions and on the use of the so-called \emph{Renewal Theory}, a well-known mathematical stochastic theory (for the reader's convenience, we provide a brief account of this theory in Supp. Mat. \ref{appendix:RT}). It is classically used in foraging theory but has never been used, to the best of our knowledge, to derive functional responses and bridge the gap between behavioral ecology and population ecology. We first show how functional responses and their stochastic fluctuations can be approximated under a wide range of possible individual behaviors and interaction types. Second, we show how our model can be used to derive stochastic versions of classical functional responses, such as Holling functional responses. We then show how to derive stochastic functional responses in many different ecological contexts, through two examples: the rate of successful copulations for males in a nuptial-feeding species, and the feeding rate of predators when handling of the prey depends on its size. Finally, we apply our framework for inference through a model comparison framework by reanalyzing a dataset on grey partridges \citep{bakeretal2010}. We show in particular that information can be extracted from observed fluctuations in order to estimate model parameters and improve inference, given that variability due to interactions themselves are large enough relatively to other sources of variations. 

\section{Functional responses from Renewal Theory: A general framework}

Consider a community with three species denoted by $e_x$, $e_y$ and $e_z$. We take three species for concreteness, one can of course consider more species, and one can consider genotypes, phenotypes, substrates or resources instead of species. Our goal is to determine the number of times $N\!_{\Delta}(x,y,z)$ a focal individual of species  $e_x$  successfully interacts with other individuals of a given species $e_y$ (possibly the same species), during a time span $\Delta$, with $x, y, z$ the size or density of each species in the environment.

\begin{figure}[!ht]
  \includegraphics[angle=90,scale=0.7
  ]{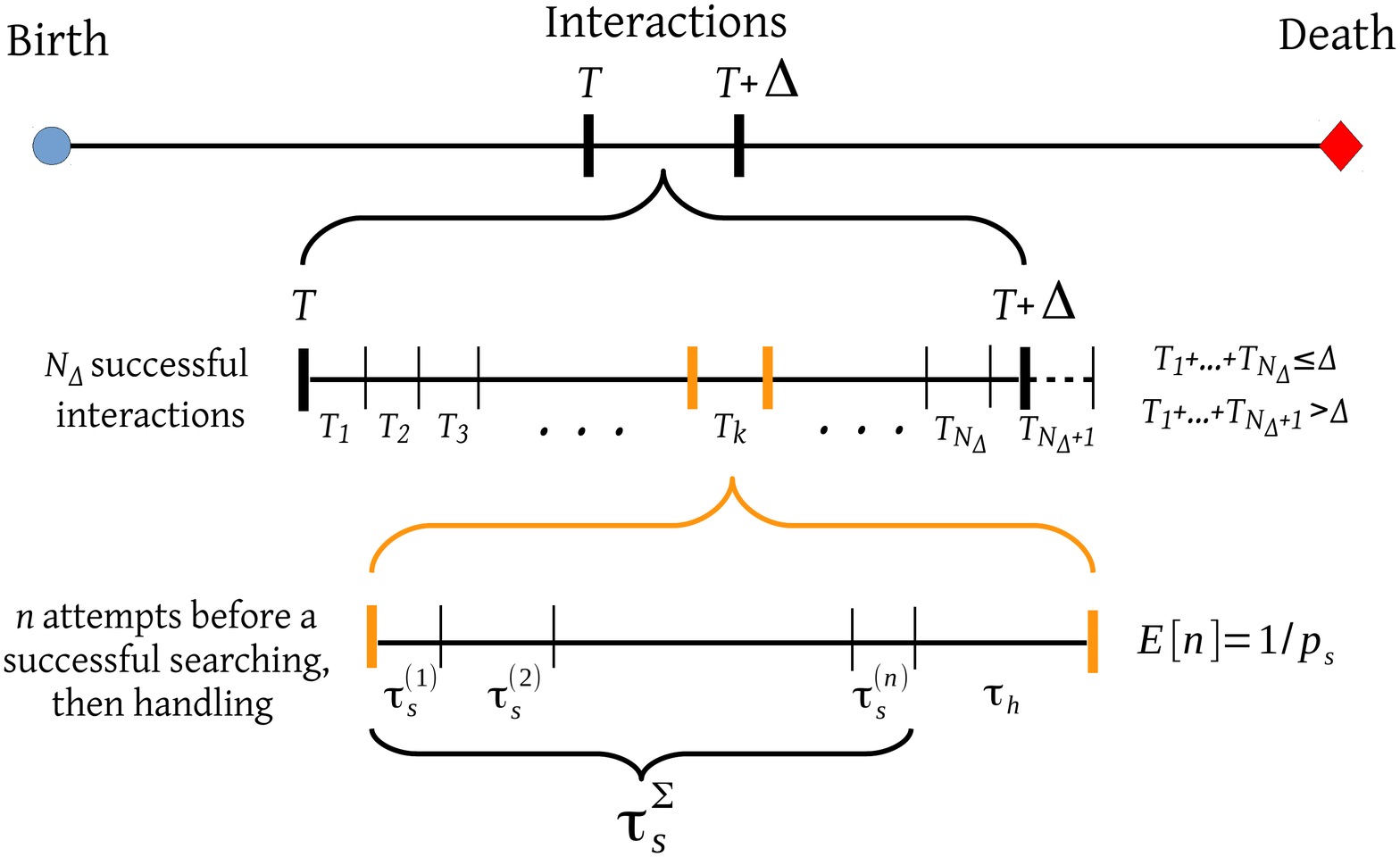}
\caption{Functional response as a decomposition of the stochastic times between two successful interactions. A time frame with duration $\Delta$ of an individual is first decomposed into $N_\Delta$ successful interactions with other individuals, for instance prey. Each successful interaction takes a random time $T_i$. Second, $T_i$ is decomposed into a sequence of necessary activities the individual must succeed in order to accomplish a successful interaction. Each activity takes a random time $\tau_a^{(j)}$ with a probability $p_a$ to succeed. As an example, two activities are necessary in the figure: searching and handling a prey. Each searching period takes a time $\tau_S^{(j)}$, has a success probability $p_s$. The individual makes $n$ searching attempts before catching a prey, resulting in a total searching time $\tau_S^\Sigma$, then handling the prey takes place which takes a time $\tau_h$ with a probability $p_h=1$ to succeed.   }\label{fig:FRschema}
\end{figure}

We assume $1 \ll N\!_{\Delta}(x,y,z) \ll x,y,z$, \textit{i.e.} i) the number of interactions is low compared to the size of the populations during $\Delta$, and ii) the number of interactions during $\Delta$ is large. In other words, it is assumed that interactions between individuals have a negligible effect on the population sizes and the time span $\Delta$ is large enough for many interactions to occur. Under these assumptions, the time between  the $k-1^{th}$ and $k^{th}$ interactions, denoted by $T_k(x,y,z)$, is a random variable whose distribution is independent of $k$, but generally depends on $x,y,z$. 
The number of interactions $N\!_{\Delta}$ during $\Delta$ is thus a random variable defined as follows (Figure \ref{fig:FRschema}):
\[
N\!_{\Delta}=k\quad \text{if}  \quad T_1+\cdots+T_k \leq \Delta < T_1+\cdots+T_k+T_{k+1}.
\]

\paragraph{Approximation of the mean and variance of a functional response.}

Functional responses, \textit{i.e.} the expected number of successful interactions per unit of time (in the more specific case of predator-prey interactions the number of prey killed by a predator per unit of time, \emph{e.g.} \cite[][]{holling1959,arditiginzburg2012, abrams2015}), can be defined as
\begin{equation}\label{def:FR}
R(x,y,z)=\frac{N\!_{\Delta}(x,y,z)}{\Delta}.
\end{equation}
Under the assumptions that the number of interactions is large (\textit{i.e.} $\Delta$ is large) but the environment does not change during this time (\textit{i.e.} variation in densities $x$, $y$ and $z$ is negligible), the mean and variance of the functional response $R(x,y,z)$ can be approximated in law according to Renewal Theory by
\begin{equation}\label{eq:LargeNumberLaw}
R(x,y,z)
\approxd\frac{1}{\E\left(T(x,y,z)\right)}+\mathscr{N}\left(0,\frac{1}{\Delta}\frac{\V\left(T(x,y,z)\right)}{\E\left(T(x,y,z)\right)^3}\right),\;\;
\end{equation} 
where $\mathscr{N}(0,\sigma^2)$ denotes a Normal distribution with mean $0$ and variance $\sigma^2$ (see Supp. Mat. \ref{appendix:RT} for mathematical details). Since all the random variables $T_k(x,y,z)$ have the same distribution, we dropped the index $k$ and simply wrote $T(x,y,z)$ inside the mean and the variance. This formula shows that knowing the expected time between two successful interactions $\E\left[T(\cdot) \right]$ and its variance $\V \left[ T(\cdot)\right]$ provides the mean of a functional response and a confidence interval. It is important to notice that the stochastic fluctuations of the functional response around its mean are only due to the interactions \emph{per se} and not to external sources: since we consider interactions between individuals as a stochastic process, it implies stochastic fluctuations. We will call this intrinsic source of variability \emph{interaction stochasticity} in order to differentiate it with possible others, such as environmental stochasticity or between individuals variability.

In order to give an explicit form of a functional response, the next step is to decompose the time $T(x,y,z)$ according to the different events taking place between two interactions.

\paragraph{Decomposing the time between two interactions.}
We suppose that the time between two interactions, namely $T(x,y,z)$, can be decomposed into a sequence of a given number of independent steps, in a definite order. We start with a simple example, and give the general version afterwards. Consider a predator $e_x$ and a prey $e_y$ (the third species  $e_z$ is ignored for simplicity). Assume that a predator has two activities:  searching a prey and handling it if its capture is successful (see Fig. \ref{fig:FRschema}). Suppose it takes a time $\tau_{h}$ to handle a prey and $\tau_{s}$ to find it. The first attempt of the predator to find a prey takes a random time
$\tau_{s}^{(1)}$. The prey is then either caught (`success') or not (`failure'). If the capture is successful, then the time of an interaction is $T(x,y)=\tau_{s}^{(1)}(x,y)+\tau_{h}(x,y)$. Otherwise, the predator has to make another attempt, which takes a new random time $\tau^{(2)}_{s}$. If there is a success, then
$T(x,y)=\tau_{s}^{(1)}(x,y)+\tau_{s}^{(2)}(x,y)+\tau_{h}(x,y)$, and so on and so forth. The reader can guess that the number of steps is given by the number of trials needed to get the first success. It is well-known that this number is a random variable which follows a geometric law whose parameter is the probability of a success.

The decomposition of the time between two interactions can be generalized and applied to many different ecological contexts, for any number of activities. Denote by $\A$ the (finite) set of possible activities and by $|\A|$ its cardinality, \textit{i.e.} the number of necessary activities in order to perform a new successful interaction. In the previous example, we had $\A=\{``\text{handling}", ``\text{searching}"\}$ and $|\A|=2$. 
Let $\tau_a(x,y,z)$ be the random time needed to carry out a given activity $a\in \A$, {\em e.g.}, $a=``\text{searching}"$. Let $p_a(x,y,z)$ be the probability that this activity ends up successfully, and $G_a(x,y,z)$ the random number of attempts the focal individual must perform before succeeding activity $a$.  $G_a(x,y,z)$ is a random variable following a geometric distribution with parameter $p_a(\cdot)$. Then we can write (the notation $(x,y,z)$ is dropped hereafter not to overburden the notations, but the reader should keep in mind that all random variables generally depend on $x,y,z$):
\[
T=\sum_{a=1}^{|\A|} \tau^\Sigma_a
\]
where $\tau^\Sigma_a$ is the total time spent for activity $a$ to end up successfully, given by
\[
\tau^\Sigma_a= \sum_{i=1}^{G_a} \tau^{(i)}_a,
\]
where, the random variables $\tau^{(i)}_a$, $i=1,2,3,\ldots$, are independent and have the same distribution as $\tau_a$ (note that $G_a$ appears as a superscript because it is the random number of attempts before succeeding activity $a$).

We can compute the mean and variance of the time span $T$ between two interactions, and we find (see Supp. Mat. \ref{app-rappels})
\begin{equation}
\label{eq:MoreGeneralET-1}
\E(T) = \sum_{a=1}^{|\A|}  \frac{\E(\tau_a)}{p_a}
\end{equation}
and
\begin{align}
\label{eq:MoreGeneralET-2}
\MoveEqLeft[4] \V\left(T\right) =  \sum_{a=1}^{|\A|} \left( \frac{\V\left(\tau_a \right)}{p_a}+\frac{1-p_a}{p_a^2}\, \E\left(\tau_a\right)^2 \right).
\end{align}



We finally obtain an approximation of the functional response 
by plugging Eqs. \eqref{eq:MoreGeneralET-1} and \eqref{eq:MoreGeneralET-2} into
Eq. \eqref{eq:LargeNumberLaw}. To achieve this, one needs to know $\tau_a(x,y,z)$  and $p_a(x,y,z)$ for all $a\in \A$,
that is, the time taken by every activity and their probability of success, both generally depending on the species density in the community.
These quantities can be determined empirically, or explicitly specified in various ecological contexts. We give examples and applications in the following.

\section{Examples of functional responses with searching and handling times}

In this section, we show how stochastic versions of classical functional responses can be derived assuming simple processes at the level of the individuals for searching and handling activities. 

\paragraph{A general form of the functional response with searching and handling times.}
Our aim is to determine the expectation and variance of the number of interactions between a focal individual $e_x$ and individuals $e_y$ during a time $\Delta$. We want to take into account possible perturbations due to interactions between the focal individual $e_x$ with individuals of a third species $e_z$. We suppose that the time between two interactions $T$ in a given environment $(x,y,z)$ can be decomposed into a searching time $\tau_s$ and a handling time $\tau_h$, that handling always succeeds ($p_h=1$) and catching an individual $e_y$  has a probability $p_s$ to succeed. If searching fails, the focal individual starts a new searching phase or stop searching if the total time $\Delta$ is reached (Figure \ref{fig:FRschema}). From Eqs. \eqref{eq:MoreGeneralET-1} and \eqref{eq:MoreGeneralET-2}, we can compute the expectation and variance of the time separating two interactions between the focal $e_x$ individual and $e_y$ individuals:

\begin{equation}\label{eq:GeneralFRmoments-E}
\E(T) =\frac{1}{p_s}\, \E(\tau_s)+\E(\tau_h)
\end{equation}
and
\begin{align}\label{eq:GeneralFRmoments-V}
\MoveEqLeft \V\left(T\right) =  \frac{\V\left(\tau_s\right)}{p_s}+\frac{1-p_s}{p_s^2}\,
\E\left(\tau_s\right)^2 + \V\left(\tau_h\right)
\end{align} 
which give an approximation of the stochastic functional response $R$ (Eq. \eqref{eq:LargeNumberLaw}). 
 In order to give explicit expression of the mean and variance of the time between two interactions $\E(T)$ and $\V(T)$, Eqs. \ref{eq:GeneralFRmoments-E} and \ref{eq:GeneralFRmoments-V} show that we now need to specify the mean and variance of the time taken to find an individual $e_y$ ($\E(\tau_s)$ and $\V(\tau_s)$), the probability to effectively interact with an individual once found ($p_s$), and the time taken for handling the interaction ($\E(\tau_h)$ and $\V(\tau_h)$). In other words, we need to explicitly specify how individuals move into space and how interactions take place. In the following, we derive functional responses under explicit assumptions regarding foraging and handling times. We first detail a simple case with assumptions which are classically made to obtain a Holling type II functional response, \emph{i.e.} constant handling time, a single type of prey, and foraging time as a function of prey densities in the environment. Second, we give a generalized model, in a $d$-dimensions space which can include many different classical functional responses.

\begin{figure}[!h]
\includegraphics[angle=0,scale=0.9]{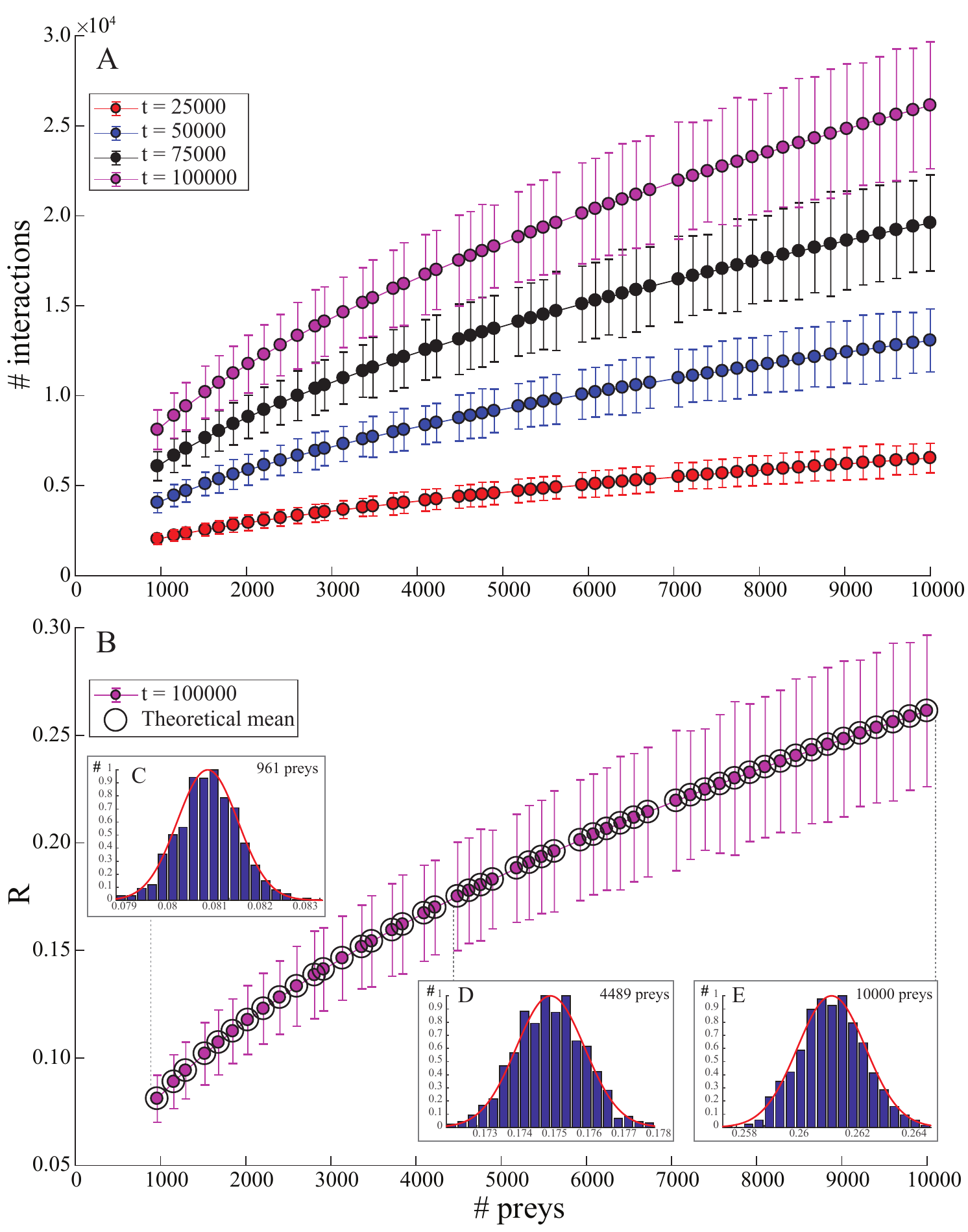}
\caption{{\small Comparison between expected and observed functional responses in individual-based simulations. Model assumptions: 2D space, uniform repartition of prey on a square lattice, constant handling time and movement speed, straight movement of the focal predator to prey, the environment is regenerated after each interaction. Interactions are counted until a maximal time is reached. Results are shown for 1000 independent runs for each parameter set. A: Dots show the observed mean number of interactions in simulations. Plain curves are the expected number of interaction, bars show the observed variance. B: observed and expected (from Eq. \ref{eq:Holling2D}) functional response (number of interactions relative to time). C, D and E: Histograms showing the empirical distribution observed in simulations for various number of prey in the environment, and plain curve shows the expected distribution (from Eq. \ref{eq:Holling2D}).}} \label{fig:Holling2D}
\end{figure}

\paragraph{A simple case: foraging in a 2D space with handling.}
We assume a focal individual $e_x$ foraging in a 2D space of size $L^2$, where $y$ individuals $e_y$ are uniformly distributed on a square lattice, such as the distance between the two nearest $e_y$ individuals is $L/(\sqrt{y}-1)$. The location of the individual $e_x$ is randomly chosen in the 2D space and we want to calculate the expectation and variance of the distance $D(y)$ to the closest $e_y$ individual. The horizontal $u_1$ and vertical distances $u_2$ between $e_x$ and $e_y$ follow uniform distributions in the range $\left[0, \delta_2 \right]$ with $\delta_2=L/2(\sqrt{y} -1)$, which gives
 \begin{align*}
   \E\left( D(y)\right)&=\frac{1}{\delta_2^2}\int_0^{\delta_2}\int_0^{\delta_2}\sqrt{u_1^2 + u_2^2} \ \text{d}u_1 \text{d}u_2 = C_ 2^E \ \frac{L}{2 \ (\sqrt{y}-1)},\\
   \E\left(D(y)^2 \right) &=\frac{1}{\delta_2^2}\int_0^{\delta_2}\int_0^{\delta_2} \left(u_1^2 + u_2^2\right) \ \text{d}u_1 \text{d}u_2 = C_2^V\left(\frac{L}{2\ (\sqrt{y}-1)}\right)^2,
 \end{align*}
 where $C_2^E$ and $C_2^V$ are two constants (explicit values are given in Supp. Mat. \ref{app-movement} ). It is assumed that the focal individual $e_x$ has a perfect knowledge of the spatial distribution of individuals $e_y$ and goes to the nearest $e_y$ individual following a straight line at speed $v$. Hence, the expectation and variance of the searching time are  $\E\left( \tau_s \right)=\E\left( D(y)\right) / v$ and  $\V\left( \tau_s \right)=\V\left( D(y)\right) / v^2$. Assuming that searching always succeeds ($p_s=1$) and that handling time is a constant $c_y$ ($\V\left( \tau_h \right)=0$), we finally get from Eqs. \ref{eq:GeneralFRmoments-E} and \ref{eq:GeneralFRmoments-V} the expectation and variance of the functional response $R$ 
\begin{equation}\label{eq:Holling2D}
 \begin{aligned}
   \E\left(R\right)&= \frac{\sqrt{y}-1}{c_y (\sqrt{y}-1) + C_2^E \lambda},\\
   \V\left(R\right)&=\frac{1}{\Delta} \frac{\left(C_2^V -{C^E_2}^2 \right) \lambda^2 (\sqrt{y}-1)}{\left( c_y (\sqrt{y}-1) + C_2^E \lambda\right)^3},
 \end{aligned}
\end{equation}
 with $\lambda=L/2v$, the scaled size of the environment (the size of the environment $L/2$ relative to foraging speed $v$). Predictions provided by Eqs. \ref{eq:Holling2D} are in very good agreement with individual-based simulations of the present model (Figure \ref{fig:Holling2D}).  

The functional response given by Eq. \ref{eq:Holling2D} is a saturating function of the density $y$ which looks like a Holling II functional response. Classically, the form of the Holling II functional response is justified by two mechanisms: a searching time depending on the density of the prey, and a constant handling time \citep{holling1959}. This is thus not surprising that we recover a functional response close to Holling II here. However, the exact form differs: we obtain a function of $\sqrt{y}$ instead of $y$. This illustrates that being explicit about how individuals forage into the environment, with specific justifications about the mechanisms underlying interactions between individuals, can give rise to alternative functional responses (other alternatives are given in the following section). This also shows that adopting a bottom-up approach allows to estimate how variable is the number of interactions in a given ecological context due to intrinsic interaction stochasticity. In the present case, the variance of the functional response decreases with $y$.\\
 
\paragraph{Foraging and handling one or two species in a $d$-dimensions space.}
 We give here a generalization of the framework with handling and foraging in $d$-dimensions. The individual $e_x$ now forages for two possible species $e_y$ or $e_z$, with constant handling times $c_y$ and $c_z$. We make similar assumptions than in the previous section, but we introduce the parameters $\alpha$, denoting a possible preference of the species $e_x$ for species $e_z$, and $\beta$ denoting a different availability or vulnerability of species $e_z$ (\emph{e.g.} if $\beta>1$, $e_z$ is easier to be detected relatively to $e_y$, see Supp. Mat. \ref{app-movement} for details). We will moreover compare two different movements followed by the focal $e_x$ individual: a straight line to the nearest individuals (as in the previous section) or a Brownian motion. We want to calculate the time between two interactions between $e_x$ and $e_y$ which depends on: i) the number of occurrence where $e_x$ interacts with $e_z$ instead of $e_y$, which follows a geometric distribution with probability of success $y/(y+\alpha z)$; ii) the two first moments of the time taken to reach a given species in a $d$-dimensions space $\theta^E_d(w)$ and $\theta^V_d(w)$, respectively (see  Supp. Mat. \ref{app-movement} for details); iii) the handling times $c_y$ and $c_z$. Finally, the time between two interactions between a focal individual $e_x$ and individuals $e_y$ are (Supp. Mat. \ref{app-movement}):  

\begin{equation}\label{eq:RFd}
\begin{aligned}
\E\left(T\right)&=\left(\frac{y+\alpha z}{y}-1\right)\left( \theta^E_d(\beta z)+c_z\right)+ \theta^E_d(y)+c_y,\\
\V\left(T\right) &= \theta^V_d(y)-\theta^E_d(y)^2+\frac{z \alpha}{y}\left( \theta^V_d(\beta z)-\theta^E_d(\beta z)^2\right)
\end{aligned}
\end{equation}

in the case of a straight movement to the closest patch and 
\begin{equation}\label{eq:RFbm}
\begin{aligned}
\E\left(T\right) &=\left(\frac{y+\alpha \, z}{y}-1\right)\left( \theta^E_{\mathrm{{\scriptscriptstyle BM}}}(\beta z)+c_z\right)+
 \theta^E_{\mathrm{{\scriptscriptstyle BM}}}(y) + c_y,\\
\V\left(T\right) &= \theta^V_{\mathrm{{\scriptscriptstyle BM}}}(y)-\theta^E_{\mathrm{{\scriptscriptstyle BM}}}(y)^2+\frac{z \alpha}{y}\left( \theta^V_{\mathrm{{\scriptscriptstyle BM}}}(\beta z)-\theta^E_{\mathrm{{\scriptscriptstyle BM}}}(\beta z)^2\right)
\end{aligned}
\end{equation}
in the case of a Brownian motion.

With appropriate assumptions regarding underlying mechanisms, Eqs. \ref{eq:LargeNumberLaw}, \ref{eq:RFd} and \ref{eq:RFbm} are general enough to recover some classical functional responses and their variance. As shown in Table 1, assuming  a 1D space and direct foraging to the nearest individual yields the equations for Holling Type I, II and III, Beddington-DeAngelis or the Ratio-Dependence functional responses, given further assumptions about handling time and possible interference between species. On a side note, we were not able to find appropriate assumptions to recover the exact equation of the Holling Type III functional responses as defined in the literature: We found a function of the form $y^2/(1+y+y^2)$ instead of the classical form $y^2/(1+y^2)$. It can be due either to a lack of generality of Eq. \ref{eq:RFd}, or because there is no realistic biological assumptions which allows to recover the same exact form under our framework, or because Holling Type III is only correct when $y$ is large, \emph{i.e.} $y$ is negligible relative to $y^2$. Even if both functions have similar sigmoidal shapes, our results illustrate that deriving functional responses with a bottom-up approach highlights implicit and often hidden assumptions regarding the underlying mechanisms.

\begin{figure}[!h]
\includegraphics[angle=0,scale=0.62]{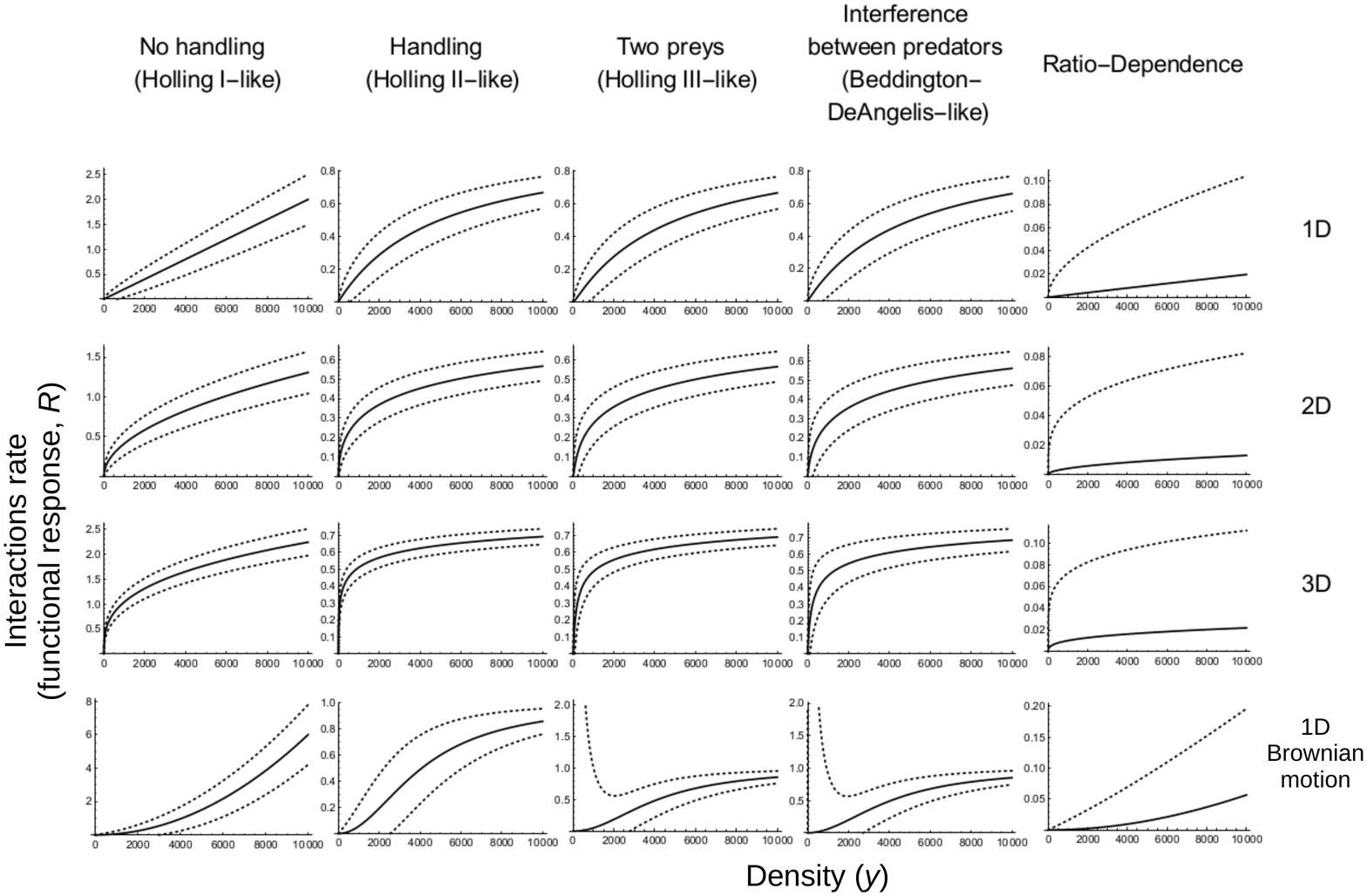}
\caption{Functional responses and their 95\% confidence interval under various underlying mechanisms regarding foraging, handling, and interaction between species.}\label{fig:FR}
\end{figure}

Interestingly, alternative underlying mechanisms can give rise to similar functional responses, in two ways. First, the exact same equation can emerge for different hypotheses. There are for instance two ways to introduce competition between predators to obtain the ratio-dependence functional response, assuming either that the probability of searching success is $p_s=1/x$ or that the searching time is proportional to the density of individuals $e_x$, $\E(\tau_s(x,y,z))=\lambda \ x/y $ and the probability of success is $p_s=1$. Second, we can obtain similar form of the functional responses with alternative foraging strategies in a given environment. Fig. \ref{fig:FR} shows that decelerating functional responses can be obtained without handling time ($c_y=0$) but with foraging in a 2D or 3D space. Sigmoidal functional responses (Holling Type III-like forms) can emerge if foraging follows a Brownian motion in a 1D space. Note that not only forms of functional responses vary with assumptions, but also their range (interactions rates have different scales even though parameter values are identical, see the $y$-axes on Fig. \ref{fig:FR}). This illustrates the relevancy of the bottom-up approach developed here. Since the same functional responses can be obtained under different hypotheses, deducing underlying mechanisms from an observed form at the macroscopic level is limited; for instance, a sigmoidal functional response does not necessarily mean that a predator is able to learn. Instead, we propose to modelize interactions at the microscopic level (the level of the individuals) in order to get corresponding functional responses.

Another merit to adopt a bottom-up approach is that the variance of the functional responses can be estimated (Tab. 1 and Fig.\ref{fig:FR}). The intensity of stochastic fluctuations of the functional responses, depicted by the 95\% confidence interval on Fig. \ref{fig:FR}, depends on the supposed underlying mechanisms (\emph{e.g.} the large fluctuations in the Ratio-Dependence functional response are due to the probability of searching success of the form $1/x$), and on the density of species (\emph{e.g.} fluctuations increase linearly with $y$ when there is no handling). Also note that even if, as shown before, we can recover the same mean functional response with different underlying hypotheses, variance can be different: the ratio-dependence functional response shows different variance whether $p_s=1/x$ and $\E(\tau_s(x,y,z))=\lambda /y $ or  $p_s=1$ and $\E(\tau_s(x,y,z))=\lambda \ x/y $. This is noteworthy because it shows that, in order to make inference from data, both the mean and the variance of the number of interactions per unit of time can be used to discriminate between concurrent functional responses models.

\begin{table}
  \footnotesize
  \centering
  \begin{tabular}{c|cccc}\label{tab:classicalRF}
   Functional & Handling  & Searching & Interaction & Mean\\
   Response & time  & probability & with $e_z$ & \& Variance ($\times \Delta^{-1}$) \\
    $R(.)$ & $c_y$& $p_s$& &  \\
   \hline \hline
   No handling & $0$  & $1$ & No: $z=0$ & $y \ l^{-1}$ 
   \\
   2 species & & & &  \\
   (Holling Type I)& & & & $y \ \left( 3 l \right)^{-1}$\\
   \hline
   With Handling &$>0$ & $1$ & No: $z=0$ &  $y\left[l + y c_y \right]^{-1}$ 
   \\
   2 species & & & &  \\
    (Holling Type II)& & & & $y l^2 \left[3\left(l+y c_y \right)\right]^{-3}$\\
   \hline
   With Handling  & $>0$ & $1$ & Yes: $z>0$ & $y^2 \left[\alpha z (c_z-l y )+ y^2 c_y   \right]^{-1}$
   \\
   3 species & & &  $c_z\rightarrow c_z/y$  &  \\
   (Holling Type III)& & & $\beta=0$ &$y^5 z \alpha l^2\left[3 \left(\alpha z \left(c_z-l y\right)+y^2 c_y\right)\right]^{-3}$\\
   \hline
   With Handling  & $>0$& $1$ & Yes : $z=x$  & $ y \left[ 2l + c_y y + c_x \alpha x\right]^{-1}$
   \\
   2 species & & & $\alpha=\beta$ & \\
   Predators Interference & & & &  $ y l^2 \left( y+\alpha x\right) \left[3 \alpha x \left(2l + c_y y + c_x \alpha x \right) \right]^{-3}$\\
   (Beddington-DeAngelis)& & & & \\
   \hline
   With Handling        & $>0$ & $x^{-1}$ & No: $z=0$ & $y/x \left[ l+c_y y/x\right]^{-1} $
   \\
      2 species        & & & & \\
   Predators Competition & & & &$y/x \ l^2 \left[ l+c_y y/x \right]^{-3}$ \\
   (Ratio Dependence)& & & & \\
   \hline
 \end{tabular}
 \caption{Examples of assumptions for recovering some classical functional responses in a 1D space with direct movement to the nearest individual (from Eq. \ref{eq:RFd}). For the sake of simplicity, formulas are given under the assumption that $y>>1$, $\alpha z>>1$ and $\beta z>>1$, and denoting $l=\lambda/2$. ($\times \Delta^{-1}$) means that the variance is scaled with the parameter $\Delta$ (see Eq. \ref{eq:LargeNumberLaw})}
 \end{table}

\section{Build the $<${\em insert your name}$>$ functional response}
In addition to recover classical functional responses, we argue that our theoretical framework is general enough to be applied to many different ecological contexts. In this section, we give two detailed examples inspired from the empirical literature. This illustrates how anyone with a specific question can derive functional responses and its stochastic fluctuations from basic knowledge about individual traits and behaviors, following a bottom-up approach.

\begin{figure}[!h]
\includegraphics[scale=0.8]{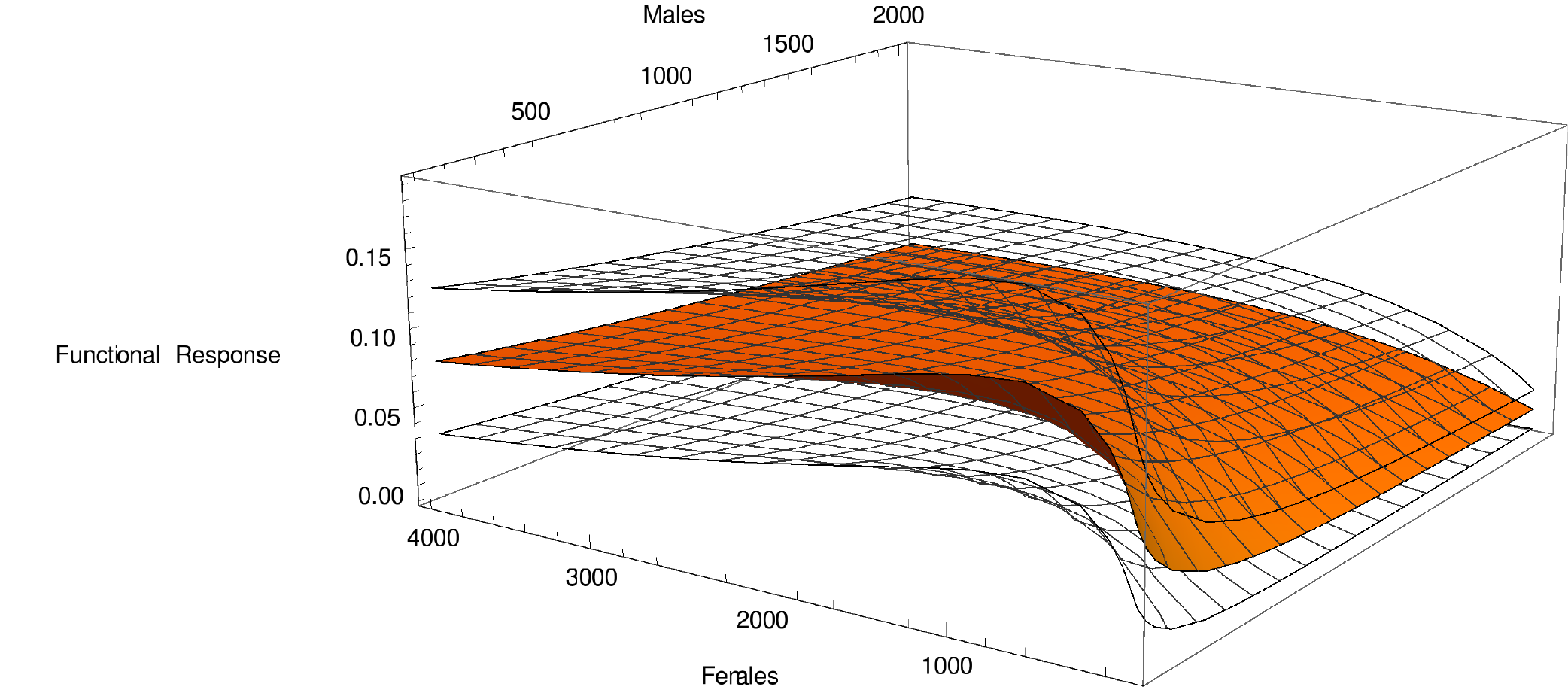}
\caption{Functional response (number of successful copulations for a male in a given time) in a nuptial-feeding species. In order to successfully mate with a female, the focal male is supposed first to catch a prey, handle it, find a free female, court it and copulate with it. Space is supposed to be 1-dimensional and the focal male to move to the nearest gift and the nearest female. The focal male is supposed to be in competition with other males and females to catch a gift, and to be in competition only with males to find a female. Handling times and court times are supposed to be constant. Parameters: Time frame duration $\Delta=100$; courtship time and probability of courtship success $C_C=1$, $p_C= 0.9$; scaled size of space $\lambda = 10$, gifts density $z= 5000$, probability of copulation success $p_R=0.5$, handling time of the gift $C_H= 1$.}\label{fig:nuptialgiving}
\end{figure}

\paragraph{Nuptial feeding: how many successful matings for a male?}
In many species such as insects, spiders or birds, males bring food to females in order to increase their probability to mate \citep{vahed1998, galvansanz2011, kleinetal2014}. Let us imagine that five successive steps must be fulfilled by a male to mate with one female: find a free a gift (\emph{e.g.} a prey), handle it, find a free female, court it and copulate. Let denote $\tau_{S_G}$, $p_{S_G}$ and $\tau_{H_G}$ respectively the time taken and probability to successfully search for a free gift (\emph{i.e.} not already handled by another male or female), and to handle it;  $\tau_{S_F}$, $p_{S_F}$ and $\tau_{H_C}$, $p_{H_C}$ the time taken and the probability to successfully search for a female, and to court it; finally let $p_R$ be the probability to successfully copulate with it. Applying Eqs. \eqref{eq:MoreGeneralET-1} and \eqref{eq:MoreGeneralET-2} gives the expectation and the variance of the time taken for a male to successfully mate with a female:

\begin{align*}
& \E\left(T\right)=\frac{1}{p_R}\left( \frac{\E\left(\tau_{S_G} \right)}{p_{S_G}} +\E\left(\tau_{H_G} \right)+\frac{\E\left(\tau_{S_F} \right)}{p_{S_F}}+ \frac{\E\left(\tau_{C} \right)}{p_{C}} \right)\\
& 
\V\left(T\right)=\frac{1-p_R}{p_R^2}\left( \frac{\E\left(\tau_{S_G} \right)}{p_{S_G}} +\E\left(\tau_{H_G} \right)+\frac{\E\left(\tau_{S_F} \right)}{p_{S_F}}+ \frac{\E\left(\tau_{C} \right)}{p_{C}} \right)^2\\
&\quad\qquad +\frac{1}{p_R}\left( \frac{1-p_C}{p_{C}^2}\, \E\left(\tau_{C} \right)^2+\frac{1-p_{SG}}{p_{SG}^2}\,
\E\left(\tau_{SG} \right)^2+\frac{1-p_{SF}}{p_{SF}^2}\, \E\left(\tau_{SF} \right)^2 \right.\\
&\quad\qquad\left.+\frac{\V\left(\tau_{S_G} \right)}{p_{S_G}} +\frac{\V\left(\tau_{S_F} \right)}{p_{S_F}}\right).
\end{align*}  
Assuming male-male and male-female competition for finding prey, the probability to successfully find a free gift by a focal male can be supposed as $p_{S_G}=1/(m+f)$, with $m$ and $f$ the male and female densities, respectively. The form given to $p_{S_G}$ is arbitrary and, assuming that all females and males have equal chance to find prey, it only reflects that the higher the density of competitor males and females, the more difficult for the focal male to find a gift. Note that the time spent by the focal male to find a free gift also depends on prey density through $\tau_{S_G}$. Similarly, we assumed that males compete for finding a free female, with equal chance, which gives the probability to successfully find a female to court is $p_{S_F}=1/m$. Handling and court times, and the probability to copulate are assumed constant (respectively equal to $C_{G}$, $C_C$ and $p_R=p$). An approximation of the expectation and variance of the number of successful matings by a male can then be obtained using Eq. \ref{eq:LargeNumberLaw}. Figure \ref{fig:nuptialgiving} shows that the variance of the number of successful copulations decreases when the number of males and females decreases, and that the expected number of successful copulations has a non-monotonous variation with the density of females. Such an approach can be useful for studying the evolution of interaction behavior in the context of nuptial-feeding behaviors \citep{galvansanz2011}. 

\paragraph{Trait-dependent interaction times: prey size and functional responses.} 
While most functional responses make the implicit assumption that interaction rates only depend on the number or density of individuals, many investigations suggest that interaction rates might also depend on individual traits and their distribution in a population, especially body mass or size \citep[\emph{e.g.}][]{aljetlawietal2004, delongetal2012}. For instance, \cite{vucicpesticetal2010} showed that functional responses generally depend on the ratio between predator and prey masses, and \cite{gonzalezsuarezetal2011} showed that predators prefer prey with a particular body mass. We show here that our theoretical framework can be used to derive functional responses dependent on quantitative traits.

As an illustration, we will focus on the derivation of handling times accounting for body mass. We neglect searching times for the sake of simplicity, but it might be relevant to include the effect of size on the time and probability to find and catch a prey: larger prey can be more easily detected or caught. The effect of prey size on searching time can for instance be taken into account into the spatial scale $L$ introduced before. Let us note $s$ the size of the prey and $m(s)$ its distribution, supposed following a truncated normal distribution defined on $[0,+\infty]$, with mean $\mu_s$ and standard deviation $\sigma_s$. $\pi(s)$ is defined as the density probability that a predator catches a prey of size $s$. Finally, we suppose that the expected handling time conditional on the prey size $s$ is $\E(\tau_h | s) = t(s)$. The expectation and variance of handling times are thus
\begin{align*}
\E(\tau_h)&=\int_0^\infty \pi(s)\, t(s) \dd s,\\
\V(\tau_h)&=\int_0^\infty \pi(s)\, t^2(s) \dd s - \big(\E(\tau_h)\big)^2.
\end{align*}

\begin{figure}[!h]
\includegraphics[scale=1]{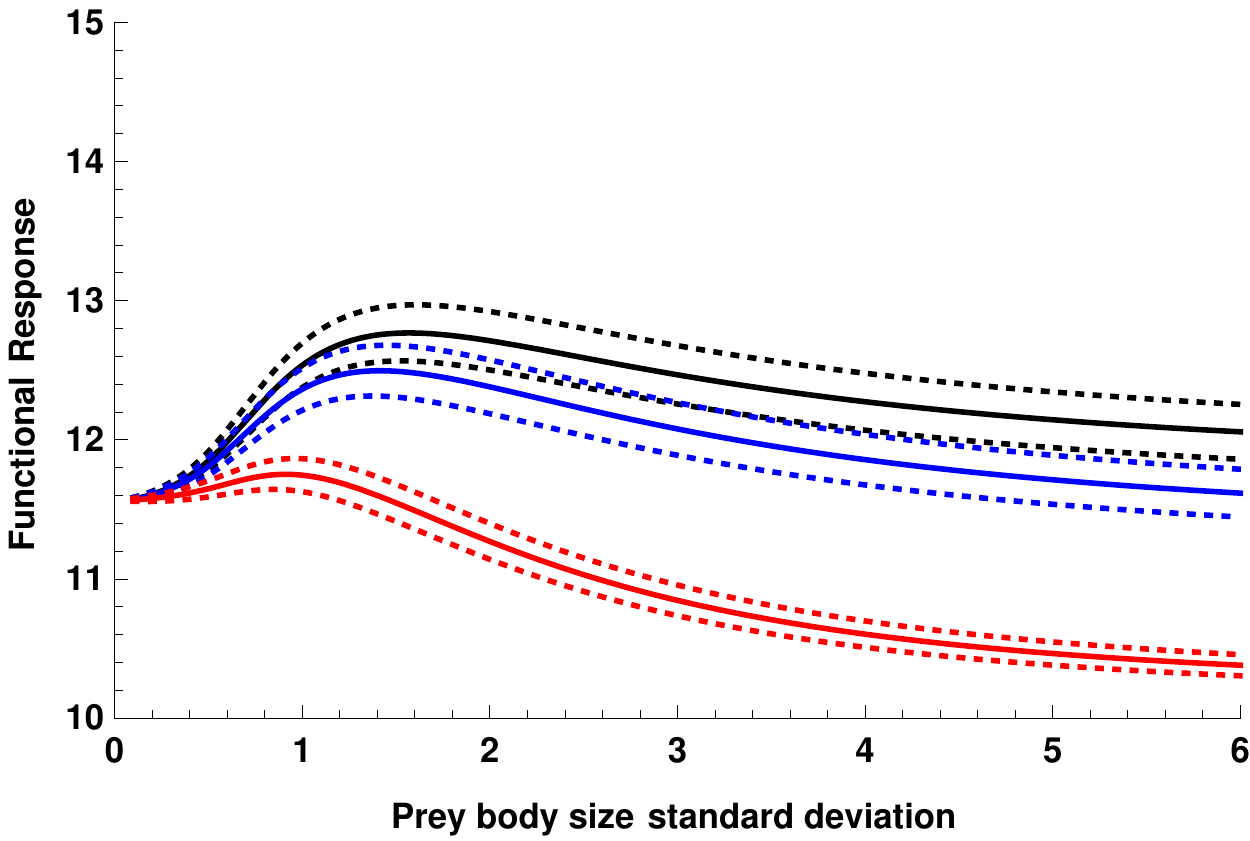}
\caption{Size dependent functional response. Handling time is assumed to depend on the prey body size. The body size and the preference of the predator are assumed to follow positive truncated normal distributions. Plain lines show the expectation of the functional response, and the dotted lines show the 95\% confidence interval. Colours show functional responses for different most preferred size by the predator: black, $\mu_p=0.1$; blue, $\mu_p=2$; red, $\mu_p=10$. Other parameters: time frame duration $\Delta=100$, mean prey size $\mu_s=2$, standard deviation of the predator preference $\sigma_P=5$, maximum handling time $T_{max}= 0.1$. The searching time is assumed to be independent to the prey size.}\label{fig:allometry}
\end{figure}

We can now be more specific in order to give explicit expression of the functional response. Let $P(s)$ be the preference of the predator for prey with size $s$, assumed to follow a truncated Gaussian function defined on $[0,+\infty]$, with $\mu_P$ the most preferred size and $\sigma_P$ its tolerance. The probability to catch a prey of size $s$ is therefore $\pi(s)=m(s) P(s) ds / \int m(u) P(u) du$. We can reasonably assume that handling time is a bounded increasing function of prey size $s$, such as $\E(\tau_h|s) = \tau_{\text{max}}(1-e^{-s})$. Finally, using Eq. \eqref{eq:LargeNumberLaw}, an approximation of the functional response and its variance can be obtained (Fig. \ref{fig:allometry}). Figure \ref{fig:allometry} shows how the functional response and its variance change as a function of the variance of the body size of the prey $\mu_s$. This illustrates that functional responses not only depend on the average body size, as shown several times \citep[e.g.][]{andresenvandermeer2010, toscanoetal2014}, but also on the variance of prey body size, possibly giving non-monotonous relationships.

\section{Functional responses inference: A model comparison framework }

Many works compare different functional responses models with likelihood ratio test or the Akaike's Information Criterion in order to determine which one fits the best \citep[\emph{e.g.}][]{skalskigilliam2001, bakeretal2010, gonzalezsuarezetal2011, hammilletal2010}. However, it is generally difficult to discriminate between alternative functional responses \citep{skalskigilliam2001,gonzalezsuarezetal2011}. A limit of these approaches is that the variance of the functional responses is directly estimated from the data, while variance depends on interactions themselves, as shown by our framework (Eq. \ref{eq:LargeNumberLaw}). Here, we reanalyze a dataset from \cite{bakeretal2010} to estimate the feeding rate of grey partridges on seeds in a controlled experiment. \cite{bakeretal2010} especially aimed at testing whether vigilant behaviors significantly affect the feeding rate. They compared two models using the AIC method: Holling type II with and without vigilant behaviors. They found no statistical difference between models with or without vigilance, and concluded that vigilant behaviors do not affect feeding rate. Here, we first derive a functional response with vigilant behaviors using our bottom-up approach. Second, we use a likelihood approach to estimate the parameters and test whether vigilance significantly affects grey partridges feeding rates (see Supp. Mat. \ref{app-Baker-general} for details).

\begin{figure}[!h]
\includegraphics[scale=1.2]{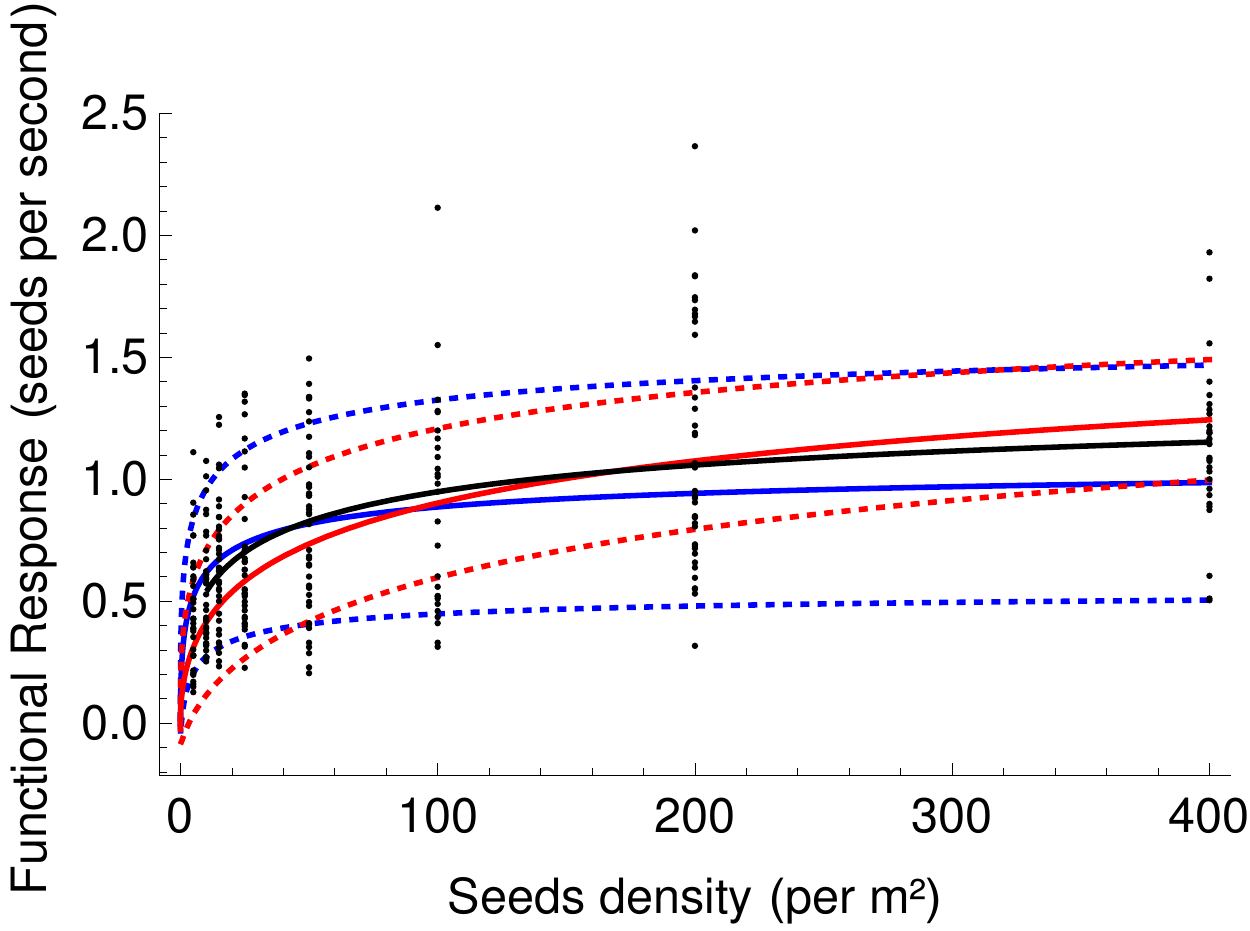}
\caption{Functional response inferred from data of the feeding rates of grey partridge (number of seeds eaten per second). Dots: empirical data. Plain and dotted curves are respectively the expectation and the confidence interval of the functional responses (when applicable). In blue: with vigilance; In red: without vigilance. In black: model fitting with variance supposed independent of the parameters (see text for details).}\label{fig:vig}
\end{figure}

We supposed that the time between two successful interactions, {\em i.e.} between two eaten seeds, can be decomposed into vigilance bouts with constant duration $c_v$, occurring with probability $p_v$, searching bouts $\tau_s$, successful with probability $p_s$, and handling bouts with constant duration $c_h$. We fitted and compared models with ($p_v \neq 0$) or without  vigilance ($p_v=0$) using a maximum likelihood approach. Note that parameters of the model were estimated both from the mean and variance (since mean and variance of the functional responses are explicitly expressed as a function of the ecological parameters). The best model was estimated to be in a 2D space, with a direct movement to the nearest seed, with vigilance (Likelihood-ratio of the models with and without vigilance = 641.3). We estimated the probability of entering a vigilant bout $p_v \simeq 0.12$, which roughly corresponds to the estimation from direct observations in \cite{bakeretal2010}. A comparison of the resulting functional responses is shown on Fig. \ref{fig:vig}.

To illustrate the importance of the information included in the variance, we also fitted the models assuming a fixed variance independent of the parameters. In this case, the model without vigilance is the best (using an AIC comparison) because it has a lower number of parameters to estimate. In practice, only using the expectation of the functional response makes difficult to distinguish between different functional responses. This is however made possible using the variance. In our case, the large variability in the functional responses are better explained by the model with vigilance because the probability to enter into a vigilance bout mechanically increases variance in our model (Eq. \ref{eq:LargeNumberLaw}).

These results must yet be taken with caution. The large variation in the functional response shown in the data (Fig. \ref{fig:vig}) can be due to other sources of variability, observation errors or variability between individuals, not taken into account into the present analyses. Indeed, as shown in \cite{schroderetal2016} for a predator fish, most of the variation in the number of prey consumed by unit of time can be due to variation in the predator individual traits: searching  and handling times both decrease with predator body size. This suggests, in the case of the grey partridges treated here, that the large observed variance is not necessarily due to the probability to enter into a vigilance bout, but to between-individuals variability. In order to properly analyze such type of data and improve our ability to infer functional responses parameters, an adapted statistical framework must be developed taking into account both between- and within-individuals variability. Within-individuals variability is taken into account  by our model (Eq. \ref{eq:LargeNumberLaw}) but a mixed-effect model is needed to take into account between-individuals variability. Ideally, taking account of  between prey variability would also be needed, for instance to take into account the effect of the variability in prey body size. Classically, mixed effects models include between-individuals variability only in the mean part of the model, and not in the variance. In our model, since between-individuals variability would affect both the mean and the variance of the model (Eq. \ref{eq:LargeNumberLaw}), a specific  mixed-effect statistical framework is needed (development in progress). 

\section{Discussion}
\par {\bf Dealing with identifiability.}
The rate at which individuals interact, within or between species, is central in most ecological processes: it affects individuals' growth, birth and death, populations' and communities' dynamics, and how energy and matter flow through ecosystems. Predicting ecological dynamics and supporting management decisions depend on our understanding of how individuals interact. It is thus crucial to use functional responses models that fit the best with data and with mechanisms underlying interactions \citep{pettorellietal2015}. However, functional response models are hardly identifiable: often, different functional responses  fit well to data and inferences show no or few statistical power \citep{skalskigilliam2001,gonzalezsuarezetal2011}. This identifiability problems of functional responses sometimes even resulted in violent debates \citep{arditiginzburg2012, abrams2015}. Adopting a bottom-up approach, we show in the present paper that such an identifiability is not surprising since, on the one hand, various underlying mechanisms can give rise to similar or identical functional responses (Fig. \ref{fig:FR}). On the other hand, our results show that choosing a functional response with a phenomenological approach does not give insight about the underlying mechanisms.

\begin{figure}[!h]
\includegraphics[scale=0.65]{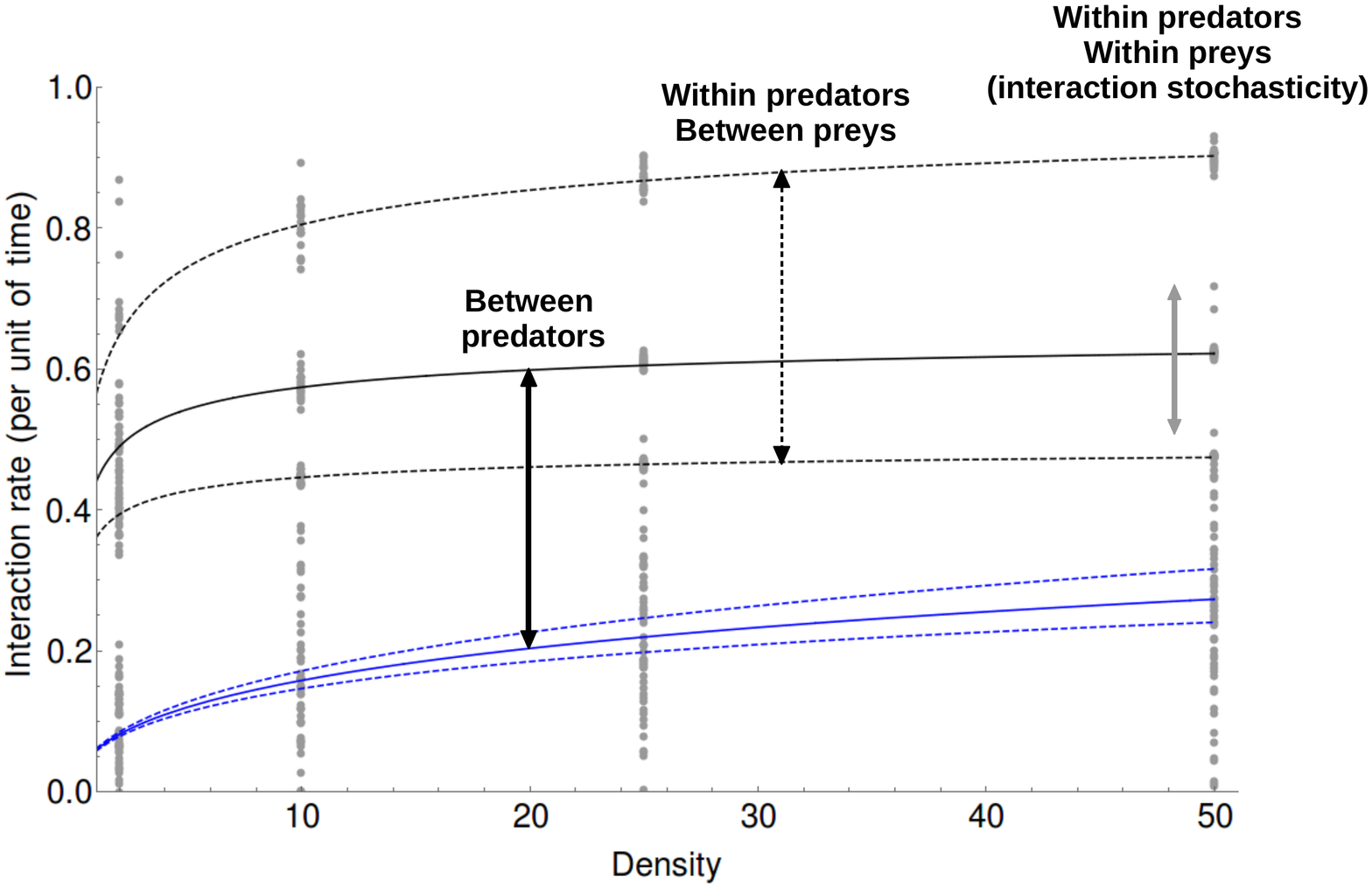}
\caption{Illustration of the three sources of variability in functional responses. Grey dots are hypothetical (simulated) measured individual number of interactions per unit of time (for instance number of prey consumed per predator). Curves represent expected functional responses under the hypothesis of a Holling Type II in a 2D space (Eq. \ref{eq:LargeNumberLaw} and \ref{eq:RFd}). Black and blue curves show variation due to between predators variability (searching rates are supposed different for each individual predator). Dashed and plain curves show variation due to prey size variability (varying handling times are supposed). Grey arrows illustrate the expected variability only due to interaction stochasticity (\emph{i.e.} for a given set of parameters: unique handling time and searching rate).}\label{fig:sources}
\end{figure}

Ignoring the various sources of variability of the interactions rates is certainly responsible in part of the difficulty to infer functional responses from data. It is now recognized that both interspecific and intraspecific variations are important in ecology \citep{bolnicketal2011, violleetal2012}, especially for functional responses \citep{pettorellietal2015}. For instance, \cite{schroderetal2016} and \cite{kalinoskianddelong2016} respectively showed that functional responses depend on predators and prey body mass. Here we show that, in addition to within and between species variability, a third source of variability should also be considered: \emph{interaction stochasticity} (Fig.\ref{fig:sources}). Since ecological processes at the level of the individuals are stochastic, the number of interactions per unit of time is random and make functional responses randomly fluctuate. We argue that the variation in the number of interactions per unit of time can have three internal sources: within species variations for both types of interacting individuals (\emph{e.g.} the predator and the prey), and \emph{interaction stochasticity} (Fig. \ref{fig:sources}). Inferring functional responses from data would need to take those three sources of variability into account. Our model would allow the development of a statistical framework adapted to this problematic since it follows a bottom-up approach, from the individual to the macroscopic level. Since our model allows to relate the variance of functional responses to ecological parameters and individual traits (Eq. \ref{eq:LargeNumberLaw}), variance in data can be used as a source of information to infer parameters (as we showed in reanalyzing \cite{bakeretal2010}'s dataset). In many cases, one can expect interaction stochasticity to be the weakest source of variability in real data. As it is illustrated in our reanalysis of the grey partridges dataset, relying on it alone is certainly not relevant for model identification. Developing a statistical framework allowing to disentangle all sources of variability and to extract information from variance is a new challenge. 

However, one of the main message of our paper is to question the relevancy of making inferences by using phenomenological functional responses: is it really important to know whether an hyperbolic Holling Type II functional response fits better the data than a sigmoidal Ginzburg-Arditi functional response \citep{hossieetal2016}? We think it is more relevant to make the link between behavioral ecology, foraging theory, population ecology and community ecology. We argue that our model can help in doing this by i) stopping searching which \emph{adhoc} functional response fits best to data, and ii) constructing one's own functional response from one's own specific case, using basic knowledge about species and their ecological contexts. Our model is general enough to adapt to many different situations, given it is possible to decompose interactions into a sequence of different activities. 

\par {\bf Assumptions and time scales.}
Many situations in natural populations or experiments are obviously expected to depart from our model. We made some choices for the sake of simplicity, without really being necessary. For instance, we supposed that individuals move in straight line or following a Brownian motion, which can be modified accordingly to a specific ecological context. However, two assumptions are inherent to our theoretical framework, and for which departures can be generally found in nature. First, we assume that the interactions can be decomposed into a sequence of activities whose duration are random, independent and identically distributed, which is a fundamental hypothesis of Renewal Theory used here. Yet, individuals are in general able to learn or sets mark in the environment in order to improve their ability to search, or change their preference for a prey or another, which violates the independence in time \citep{holling1959, yazdaniandkeller2016}. Individuals can also be clustered in space, which would for instance affect the time taken by a predator to find two succeeding prey \citep{hossieetal2016}. Population size of each interacting species can also significantly vary during the considered time frame, because of prey depletion or blooming. Second, we assume that the number of interactions and the considered time frame are large. Yet in general, experiments and observations generally involve a few dozens or hundreds of interacting individuals during a limited time frame. 

What is the impact of violating these assumptions on the resulting functional response is an open question. Performing stochastic simulations might obviously help in addressing this question. However, we think it can be addressed more generally from a theoretical point of view. Indeed, we think that it mostly relies on how do  the different time scales of the ecological processes involved in the functional response relate with each other. For instance, if individual learning occurs at a slower time scale than it consumes prey, then we can expect that learning little affects the resulting functional response. Similarly, if the time taken by a predator to move from one patch of prey to another is large enough, then it should not affect much the resulting functional response in a short time frame. If prey reproduction is fast enough, then depletion should have a low impact. Some authors proposed for example to take into account prey depletion into account \citep{rogers1972}, but the approach is still phenomenological, \emph{i.e.} different time scales are generally not considered. The relative importance of the time scales is central in ecology. It deserves particular attention in the case of functional responses in particular \citep{getz1998}, and it is a natural perspective of the current work. 

The last perspective of this work is about how individuals interactions translates into birth and death, in other words how functional responses are related to numerical responses \citep{pettorellietal2015}. Since most population and community dynamics models are systems of differential equations, functional responses and numerical responses are generally assumed to play at similar time scales (except when slow/fast processes are explicitly assumed). Some authors showed in specific cases that considering different time scales for functional responses and numerical responses \citep[\emph{e.g.}][]{getzandschreiber1998, casasandmccauley2012} can dramatically affect population dynamics and stability.  Finally, functional response should be considered as the result of stochastic ecological processes. We highlight in the present paper the phenomenon we called \emph{interaction stochasticity}, in analogy with \emph{demographic stochasticity}. It is well-known that demographic stochasticity can affect population stability \citep{landeetal2003}. How interaction stochasticity can affect population and community stability, through its effect in numerical responses, is an open question.

\section*{Author Contributions}
All authors designed the theoretical framework. SB wrote the first draft of the manuscript, and all authors contributed substantially to revisions.
\section*{Acknowledgements}
We are very grateful to David Baker for sharing its dataset on grey partridges. We want to thank Clotilde Lepers, Elisabeta Vergu and Arnaud Senlis for helpful discussions and Geoffroy Berthelot for providing simulations. SB benefited from the support by the ``H.+C.+Y.+M.+V.+D.$^\dagger$+M. Moens-Sacchettini organization for the advancement of knowledge''.
\section*{Data Accessibility}
Data are deposited on a public server (Dryad : doi:10.5061/dryad.c73bm2q)
\section*{Funding Statement}
 This work was supported by the Chair ``Mod\'elisation Math\'ematique et Biodiversit\'e'' of Veolia Environnement-Ecole Polytechnique-Museum National d'Histoire Naturelle-Fondation X and by the ANR ABIM (ANR-16-CE40-0001).

\newpage

\renewcommand\refname{References}
\bibliographystyle{unsrtnat}
\bibliography{reference}

\newpage

\section*{Figures}

\newpage

\thispagestyle{empty}

\newpage

\newpage

\newpage

\newpage

\newpage

\newpage
\appendix
\numberwithin{equation}{section}
\section{Supplementary Materials of the paper  ``Rejuvenating Functional Responses with Renewal Theory''}
by S. Billiard, V. Bansaye and J.-R. Chazottes.

\subsection{Some basics in Renewal Theory}\label{appendix:RT}

For the convenience of the reader, we state and prove the two results we are using in this paper. We refer to 
{\em e.g.} \cite{feller2} or \cite{ross} for more details. Let $T_1, T_2,\ldots$ be a sequence of independent non negative random variables. We assume that the $T_i$'s are all distributed as a non negative random variable $T$ with distribution $F$. We assume that $F(0)\neq 1$ ({\em i.e.}, the random variables are not identically zero) and $\E(T)<+\infty$. Finally, let $S_n=\sum_{i=1}^n T_i$ for $n\geq 1$. The renewal process $(N_t)_{t\geq 0}$ associated to $T_1, T_2,\ldots$ is defined by
\[
N_t=\sum_{n=1}^{+\infty} \mathds{1}_{\{S_n\leq t\}}=\sup\{n \geq 1: S_n\leq t\}.
\]
Since $\P(T\geq 0)=1$ and $\P(T=0)<1$, we have $\E(T)>0$.
By the strong law of large numbers, we know that $S_n/n$ converges to $\E(T)$ with probability one. Therefore
we deduce that $S_n\to+\infty$ with probability one. In particular, with probability one, $S_n<t$ for only a finite number of $n$'s, showing that $\P(N_t<+\infty)=1$. It follows that, for all practical purposes, we can put a `max' instead of a `sup' in the definition of $N_t$.

\noindent \textbf{Strong law of large numbers for renewal processes}. We have
\[
\frac{N_t}{t} \xrightarrow[t\to+\infty]{} \frac{1}{\E(T)}\quad \text{with probability one}.
\]
\noindent \textbf{Proof.} By the very definition of $N_t$ we have $S_{N_t}\leq t < S_{N_t+1}$, implying that
\[
\frac{S_{N_t}}{N_t}\leq \frac{t}{N_t} <\frac{S_{N_t+1}}{N_t}.
\]
Now $N_t\to+\infty$ with probability one, hence
\[
\frac{S_{N_t}}{N_t} \xrightarrow[t\to+\infty]{} \E(T)\quad\text{and}\quad 
\frac{S_{N_t+1}}{N_t+1}\frac{N_t+1}{N_t} \xrightarrow[t\to+\infty]{} \E(T).
\]

To state the second result we use, we have to assume in addition that the variance of $T$, which we denote by $\sigma^2$, is finite.

\noindent \textbf{The central limit theorem for renewal processes}. We have
\[
\frac{N_t -\frac{t}{\E(T)}}{\sigma \sqrt{\frac{t}{(\E(T))^3}}} \lawto \mathscr{N}(0,1),
\]
where $\mathscr{N}(0,1)$ is the standard Gaussian distribution (centred at $0$ and with variance equal to $1$).

\noindent \textbf{Proof.} Let $r_t:=\frac{t}{\E(T)} + y\sigma \sqrt{\frac{t}{(\E(T))^3}}$. If it is an integer, let $n_t=r_t$, if it is not an integer, let $n_t=\lceil r_t\rceil +1$. Then
\begin{align*}
\P\left(\frac{N_t -\frac{t}{\E(T)}}{\sigma \sqrt{\frac{t}{(\E(T))^3}}} <y\right)
&= \P(N_t<r_t) = \P(N_t<n_t)\\
&= \P(S_{N_t}>t) \quad \text{(because}\; \{N_t<n\}=\{S_n>t\}\text{)}\\
&= \P\left( \frac{S_{n_t}-n_t \E(T)}{\sigma \sqrt{n_t}}> \frac{t-n_t \E(T)}{\sigma \sqrt{n_t}}\right).
\end{align*}
We now use the central limit theorem for the sequence $T_1, T_2,\ldots$, and the fact that
\[
\lim_{t\to\infty} \frac{t-n_t \E(T)}{\sigma \sqrt{n_t}}=
\lim_{t\to\infty} \frac{t-r_t \E(T)}{\sigma \sqrt{r_t}}=
\lim_{t\to\infty} \frac{ -y \E(T)\sqrt{\frac{t}{(\E(T))^3}} }{ \sqrt{ \frac{t}{\E(T)}+y\sigma \sqrt{\frac{t}{(\E(T))^3}}}}=-y.
\]
It follows that, if $Z$ is a random variable distributed according to $\mathscr{N}(0,1)$, we have
\[
\lim_{t\to+\infty}\P\left(\frac{N_t -\frac{t}{\E(T)}}{\sigma \sqrt{\frac{t}{(\E(T))^3}}} <y\right) = \P(Z>-y)=\P(Z<y).
\]
This ends the proof. Loosely speaking, $N_t$ is approximately Gaussian with mean $\frac{t}{\E(T)}$ and variance
$\frac{t\sigma^2}{(\E(T))^3}$.

\vspace{0.2cm}
\noindent \textbf{Application to determine the stochastic fluctuations of the functional responses}.

To apply the two previous theorems in our context, set $t=\Delta$ and apply them for each given triplet $(x,y,z)$. More precisely, the functional response, defined as $R(x,y,z)=N\!_{\Delta}(x,y,z)/\Delta$, is a random variable and we want to determine its fluctuations around its expected value $1/\E\left(T(x,y,z)\right)$. This can also be achieved by applying Renewal Theory. Colloquially, we get that the distribution of 
\begin{equation}\label{pouic}
\frac{N\!_{\Delta}(x,y,z)}{\Delta}- \frac{1}{\E\left(T(x,y,z)\right)}
\end{equation}
is approximately a Gaussian distribution centred at $0$ with a variance equal to 
\[
\frac{1}{\Delta} \frac{\V\left(T(x,y,z)\right)}{(\E\left(T(x,y,z)\right))^3},
\]
where $\V\left(T(x,y,z)\right)=\E\left(T^2(x,y,z)\right)-\big(\E\left(T(x,y,z)\right)\big)^2$ stands for the variance of
$T(x,y,z)$. In particular, one can say that standard deviation of \eqref{pouic} is approximately
\[
\frac{1}{\sqrt{\Delta}}\, \frac{\sqrt{\V\left(T(x,y,z)\right)}}{\E\left(T(x,y,z)\right))^{3/2}}.
\]
The larger the $\Delta$, the better these approximations.
Finally, we can condense what precedes by writing  
\begin{equation}
R(x,y,z)
\approxd\frac{1}{\E\left(T(x,y,z)\right)}+\mathscr{N}\left(0,\frac{1}{\Delta}\frac{\V\left(T(x,y,z)\right)}{\E\left(T(x,y,z)\right)^3}\right),
\end{equation} 
where $\mathscr{N}(0,\sigma^2)$ is a random variable having a Gaussian distribution centred at 0 and with variance
$\sigma^2$.

\subsection{Some probability facts}\label{app-rappels}


\par{\bf Sum of i.i.d random variables $X_i$ a random number of times $Y$.}
Let $Y$ be a random variable assuming positive integer values such that $E(Y)<+\infty$. Let $X_1,X_2,\ldots$ be a sequence of independent
identically distributed random variables which are also independent of $Y$, and such that $\E(X_1)<+\infty$. Then
\[
\E\left(\sum_{i=1}^Y X_i\right)=\E(Y)\, \E(X_1).
\]
Furthermore, assume that $\V(Y)<+\infty$ and $\V(X_1)<+\infty$. Then
\[
\V\left(\sum_{i=1}^Y X_i \right)=\E(Y)\, \V(X_1) + \big(\E(X_1)\big)^2\, \V(Y).
\]

\par{\bf Expectation and variance of the time between two interactions $T$ (Eqs. \ref{eq:MoreGeneralET-1} and \ref{eq:MoreGeneralET-2}).}
Assuming that the probability of success of activity $a$ is $p_a$, then $Y_a$ the number of times activity $a$ is performed before success follows a geometric distribution with expectation and variance respectively $\E\left[ Y_a \right]=1/p_a$ and $\V\left[ (1-p_a)/p_a^2\right]$. Assuming that activity $a$ takes a time $X_a=\tau_a$, with expectation $\E\left[\tau_a \right]$ and variance $\V\left[\tau_a \right]$, which finally gives the expectation and variance of the time between two interactions $T$ in Eqs. \ref{eq:MoreGeneralET-1} and \ref{eq:MoreGeneralET-2}.

\par{\bf Expectation and variance of the time taken to search and find $e_y$ (Eqs. \ref{eq:RFd} and \ref{eq:RFbm}).} 
We want to calculate $T$ the time taken by a focal individual $e_x$ to interact with an individual $e_y$ given there are individuals $e_z$ in the environment. $Y_s$ is the the number of times the individual $e_x$ forages the environment before encountering $e_y$, hence $e_x$ interacts $Y_s-1$ times with $e_z$ before interacting with $e_y$. Assuming that $e_x$ has a probability $y/(y+\alpha z)$ to choose to searching for individual $e_y$, the expected number of time $e_x$ interacts with $e_z$ before interacting with $e_y$ is $\E\left[Y_s -1\right]=(y+\alpha y)/y-1$.  Further,in the case of a straight movement to the nearest prey, assuming that the expected time taken to search for individuals $e_z$ is $\theta^E_d(\beta z)$, that handling time of $e_z$ by $e_x$ is a constant $c_z$, that the expected time to search for individual $e_y$ is $\theta^E_d(y)$, and that handling time of $e_y$ by $e_x$ is a constant $c_y$ then the total expected time taken by interaction between $e_x$ and individuals $e_z$ is given by
\begin{align*}
\E\left[ T | Y_s \right] &= (Y_s-1) \left(\theta_d^E (\beta z) +c_z\right) + \theta_d^E (y) +c_y \\
 \implies \E\left[ T \right] &= \left( \frac{y+\alpha z}{y} -1 \right)\left(\theta_d^E (\beta z) +c_z\right)+ \theta_d^E (y) +c_y
\end{align*} which corresponds to Eq. \ref{eq:RFd}  (similar calculations gives Eq. \ref{eq:RFbm} in the case of a Brownian motion).

In order to calculate the variance of the time taken by a focal individual $e_x$ to interact with an individual $e_y$, we need to calculate the second moment of $T$. Assuming that the second moment of the time taken to search for individuals $e_z$ is $\theta^V_d(\beta z)$ and to search for individuals $e_y$ it is $\theta^V_d(y)$,  the second moment of $T$ is given by 

\begin{align*}
\E\left[T^2 | Y_s \right] &= (Y_s-1) \theta^V_d(\beta z)+ \theta^V_d(y) + 2 (Y_s-1) \theta^E_d(\beta z) \theta^E_d(y)+(N-1)(N-2) \theta^E_d(\beta z)^2\\
\implies E[T^2] &=\left( \frac{y+\alpha z}{y} -1 \right)\theta^V_d(\beta z)+2\left( \frac{y+\alpha z}{y} -1 \right)\theta^E_d(\beta z) \theta^E_d(y)\\
&+\left( \frac{y+\alpha z}{y} -1 \right)\left( \frac{y+\alpha z}{y} -2 \right)\theta^E_d(\beta z)^2 + \theta^V_d(y) 
\end{align*}

which after some calculations gives the variance of $T$ in Eq. \ref{eq:RFd}.

\subsection{Foraging in a $d$-dimension space}\label{app-movement}

The individual $e_x$ forages an environment where two species $e_y$ or $e_z$ are present with densities $y$ and $z$, respectively. The environment is a $d$-dimensions space with size $L^d$. The handling times of the individual $e_x$ with species $e_y$ and $e_z$ are assumed constant, respectively denoted by $c_y$ and $c_z$. We assume that the focal individual $e_x$ arrives randomly at a given position in space, and that it has a perfect knowledge of the environment. The parameter $\alpha$ denotes a possible preference of the species $e_x$ (if $\alpha>1$) or rejection (if $\alpha <$) for resource $e_z$ relatively to species $e_y$. Because of this possible preference, the individual $e_x$ chooses to interact next with an individual of species $e_y$ with probability $y/(y+\alpha z)$ or with an individual of species $e_z$ with probability $\alpha z/(y+\alpha z)$. Given the species with which individual $e_x$ will interact next, say $e_z$, we have to compute the time taken for $e_x$ to reach an individual $e_z$. The parameter $\beta$ denotes an availability or vulnerability of species $e_z$ relatively to $e_y$ during the foraging process (for instance species $e_z$ is easier to detect if $\beta>1$). The expected time to reach an individual $e_y$ is  denoted by $\theta^E_d(y)$, and the time taken to reach an individual $e_z$ is denoted by $\theta^E_d(\beta z)$. Two different movements are considered: following a straight line to the nearest patch or following a Brownian motion at speed $v$. Once the interaction has taken place, foraging starts again from a random position in the environment. There is thus no memory in the foraging process, which is a basic assumption of Renewal Theory used in our framework.

\subsubsection{Regular repartition in a $d$-dimensions space, movement to the nearest individuals}
The focal $e_x$ individual has a random location in the environment and moves to the nearest individual. We denote $D_d(y)$ the expected distance between the $e_x$ individual and and individual of species $e_y$ given there is a number $y$ individuals in the environment (similarly we denote $D_d(\beta z)$ the expected distance between $e_x$ and $e_z$). Assuming that $e_y$ and $e_z$ individuals have a regular distribution in a $d$-dimensions space of size $L^d$, the distance between two individuals $e_y$ is ${\delta_d(y)}=\frac{L}{2} \frac{1}{y^{1/d}-1}$, and  the distance between two individuals $e_z$ is ${\delta_d(\beta z)}=\frac{L}{2} \frac{1}{(\beta z)^{1/d}-1}$ (accounting for $\beta$, the vulnerability or availability of $e_z$). The two first moments of the distance $D_d(w)$ are given by 
\begin{align*}
\E\left(D_d(w)\right) &= \frac{1}{{\delta_d(w)^d}}\int ... \int_0^{{\delta_d(w)}} \sqrt{u_1^2+...+u_d^2}\, \dd u_1 ... \dd u_d= C^E_d \delta_d(w), \\
\E\left(D_d(w)^2\right) &=\frac{1}{{\delta_d(w)^d}} \int ... \int_0^{{\delta_d(w)}} (u_1^2+...+u_d^2) \dd u_1...\dd u_d = C^V_d {\delta_d(w)}^2,
\end{align*}
with $C^E_d$ and $C^V_d$ two constants depending on the dimension, given by
\begin{align*}
  C^E_1 &= \frac{1}{2} \ , C^V_1 = \frac{1}{3}; \\
  C^E_2 &= \frac{\sqrt 2+\log(1+\sqrt 2 )}{3} \ , C^V_2 = \frac{2}{3};\\
  C^E_3 &= \frac{6 \sqrt 3- \pi+\log(3650401+2107560\sqrt 3 )}{24},  C^V_3 =  1.\\
\end{align*}

 Given that the $e_x$  individual moves at speed $v$, the two first moments of the time taken by the $e_x$ individual to encounter an  $e_w$ individual in a $d$-dimension space are given by
\begin{align*}
  & \theta^E_d(w)=\E\left(\frac{D_d(w)}{v}\right) =C^E_d \frac{\delta_d(w)} {v},\\
& \theta^V_d(w) = \E\left(\frac{D_d(w)^2}{v^2}\right)= C^V_d  \, \frac{\delta_d(w)^2}{v^2}.
\end{align*}
Finally, given the focal individual $e_x$ has a preference $\alpha$ for $e_z$ individuals, and given that the handling time of $e_z$ individuals by $e_x$ is $c_z$, then the total expected searching time $\tau_{s,d}(x,y,z)$ of an $e_y$ individual by an $e_x$ individual in a $d$-dimensions space is given by
\begin{align}
\E\left(\tau_{s,d}(x,y,z)\right)&=\left(\frac{y+\alpha z}{y}-1\right)\left( \theta^E_d(\beta z)+c_z\right)+ \theta^E_d(y),\\
\V\left(\tau_{s,d}(x,y,z)\right) &= \theta^V_d(y)-\theta^E_d(y)^2+\frac{z \alpha}{y}\left( \theta^V_d(\beta z)-\theta^E_d(\beta z)^2\right)
\end{align}

\subsubsection{Regular repartition in a 1-dimension space, Brownian motion}

We assume as before  that $e_z$ have weights $\alpha$ and $\beta$, and that $e_y$ and $e_z$ are regularly distributed in space, with a distance $\delta_1(y)=L/(2 (y-1)$ between individuals $y$ and $\delta_1(\beta z)=L/(2 (\beta z -1)$ between individuals $e_z$. Note that we focus here on a $1$-dimension space for simplicity. Let us denote $H(l,a,b)$ the random variable representing the time taken by the $e_x$ individual moving following a Brownian motion to first hit the boundary of a segment $[a,b]$, given that its initial location $l$ is random in $ l \in[a,b]$. The Laplace transformation of $H(l,a,b)$ is
\[
\mathcal{L}(u)=\frac{\cosh\left[(b-2l+a)\sqrt{u/2}\right]}{\cosh\left[(b-a)\sqrt{u/2}\right]}. 
\]
The two first moments of $H(l,a,b)$ are respectively given by $\frac{\partial \mathcal{L}(u)}{\partial u}|_{u=0}$ and $\frac{\partial^2 \mathcal{L}(u)}{{\partial u}^2}|_{u=0}$ \citep{BorodinSalminen1996}. Hence, the two first moments of the time taken by an $e_x$ individual to reach either an individual $e_w$ are respectively (with $a$=0 and $b=\delta_1(w)$): $\theta_{\mathrm{{\scriptscriptstyle BM}}}^E(w)=  C^E_{\mathrm{{\scriptscriptstyle BM}}} \, {\delta_1(w)}^2 / v$ and $\theta_{\mathrm{{\scriptscriptstyle BM}}}^V(w)=C^V_{\mathrm{{\scriptscriptstyle BM}}} \, {\delta_1(w)}^4/v^2$ with $C^E_{\mathrm{{\scriptscriptstyle BM}}}=\frac{1}{6}$, $C^V_{\mathrm{{\scriptscriptstyle BM}}}=\frac{1}{15}$.
Finally, given a weight $\alpha$ of $e_z$ individuals an handling time of individuals $e_z$ by $e_x$ is $c_z$, the expected searching time of an $e_y$ individual by an $e_x$ individual in a $1$D space with Brownian motion is given by
\[
\E\left(\tau_{s,\mathrm{{\scriptscriptstyle BM}}}(x,y,z)\right) =\left(\frac{y+\alpha \, z}{y}-1\right)\left( \theta^E_{\mathrm{{\scriptscriptstyle BM}}}(\beta z)+c_z\right)+
 \theta^E_{\mathrm{{\scriptscriptstyle BM}}}(y),
\]
and its variance is
\begin{align*}
\V\left(\tau_{s,\mathrm{{\scriptscriptstyle BM}}}(x,y,z)\right) &= \theta^V_{\mathrm{{\scriptscriptstyle BM}}}(y)-\theta^E_{\mathrm{{\scriptscriptstyle BM}}}(y)^2+\frac{z \alpha}{y}\left( \theta^V_{\mathrm{{\scriptscriptstyle BM}}}(\beta z)-\theta^E_{\mathrm{{\scriptscriptstyle BM}}}(\beta z)^2\right)
\end{align*}

\subsection{Analysis of Baker et al. (2010)'s data}\label{app-Baker-general}

In the \cite{bakeretal2010} experiments, the behavior of grey partridges was observed and recorded in controlled conditions, with variable seeds densities. The feeding rate was calculated, {\em i.e.} the number of seeds eaten per second per individual. Other behaviors were also recorded such as the duration of vigilance bouts or handling times. The duration of observations bouts was at most of 240 seconds. \cite{bakeretal2010} derived a version of Holling Type II functional response with vigilance in a phenomenological manner and fitted it with their data. They showed no statistical difference between models with or without vigilance and concluding that vigilant behaviors do not affect the feeding rates of grey partridges. Here, we derive a functional response with vigilance from the level of the individuals (following our bottom-up approach). We first give the expectation and the variance of the feeding rate, and second we fit our models on \cite{bakeretal2010}'s data following a maximum likelihood approach, using both the mean and the variance of the functional response. 

\subsubsection{Derivation of a functional response with vigilance}\label{app-Baker}
We derive the functional response with vigilance by decomposing the time $T$ between two successful interactions, {\em i.e.}, two eaten seeds, in three steps. First, we suppose that the time between two eaten seeds is decomposed into a time taken to approach a seed and a time to handle it, respectively denoted by $\tau_a$ and $\tau_h$. Supposing that the handling time is fixed $\E(\tau_h)=c_h$, that handling and approaching are always successful, we have
\begin{align}
\E(T)&=\E(\tau_a)+\E(\tau_h)=\E(\tau_a)+c_h\\
\V(T)&=\V(\tau_a)+\V(\tau_h)=\V(\tau_a).
\end{align}

which gives an approximation of the stochastic functional response as given by Eq. \eqref{eq:LargeNumberLaw}.
Second, we suppose that the time taken to approach a seed is decomposed into times to successfully foraging for a seed, denoted by $\tau_f$ which succeeds with probability $p_f$. We thus have

\begin{align*}
\E(\tau_a) &=\frac{\E(\tau_f)}{p_f}\\
\V(\tau_a) &=\frac{\V(\tau_f)}{p_f}+\frac{1-p_f}{p_f^2}\, \E(\tau_f)^2.
\end{align*}

Third, we suppose that the time taken by successful foraging can be decomposed into a succession of vigilance bouts, with duration $\tau_v$ and searching time denoted by $\tau_s$, such as $\tau_f=\tau_v+\tau_v+\cdots+\tau_s$. We denote $p_v$ the probability that the animal enters in a vigilance bout, hence the expected number of vigilance bouts before encountering a seed is $1/(1-p_v)-1$. Finally, assuming that the vigilance times are fixed, $\E(\tau_v)=c_v$ and $\V(\tau_v)=0$, we have

\begin{align*}
\E(\tau_f)&=\left(\frac{1}{1-p_v}-1\right)\E(\tau_v)+\E(\tau_s)=\left(\frac{1}{1-p_v}-1\right)c_v+\E(\tau_s)\\
\V(\tau_f)&=\left(\frac{1}{1-p_v}-1\right)\V(\tau_v)+\frac{p_v}{(1-p_v)^2}\, \E(\tau_v)^2 + \V(\tau_s)\\
&=\frac{p_v}{(1-p_v)^2}c_v^2 + \V(\tau_s).
\end{align*}

The last step consists in determining mean and variance of the searching times, which depends on the movement of the individual and the dimension of space and the density of seeds $y$, as shown in the main text (see Eqs. \eqref{eq:RFd} and \eqref{eq:RFbm}). 

\subsubsection{Likelihood ratio test}\label{app-LRT}

From the model given before, we can derive an approximation of the expectation and the variance of the functional response as a function of seeds density $y$ and the set of parameters $\Theta$: the time taken by a vigilance bout $c_v$ and the probability to enter in a vigilance bout $p_v$, the time taken by handling seeds $c_y$, the probability that foraging succeeds $p_f$. Our goal is to test whether vigilant behaviors significantly affect the functional response given a data set $\mathcal(D)$ from the experiments in \cite{bakeretal2010}. To answer this question, a commonly used method is to estimate the parameters set $\Theta$ of two different models, without ($p_v=0$) or with ($p_v \neq 0$) vigilance bouts, and to compare their likelihood by a likelihood ratio test since both models are nested. 

The likelihood of the functional response $R(\Theta,y)$ given the dataset $\mathcal{D}$ is defined as
\[
\ell\left( R(\Theta,y) | \mathcal{D} \right) \equiv \P \left( \mathcal{D}|R(\Theta,y)\right)
\]
where $\P(E)$ is the probability of $E$. An approximation of $\P \left( \mathcal{D}|R(\Theta,y)\right)$ is given by Eq. \eqref{eq:LargeNumberLaw} and its derivation in App. \ref{app-Baker}. The dataset $\mathcal{D}$ contains the feeding rate as calculated for a given individual in eight treatments with different densities from 5 to 400 seeds.$m^{-2}$. We looked for the maximum of $\log \ell$ for different models (1D, 2D, or 3D space with movements to the nearest and Brownian motion in a 1D space), with or without vigilance, using the newton method in the ``FindMaximum'' procedure of Mathematica 10.1 \citep{wolfram}. Parameters estimation was constrained to adequate with their definition (for instance probabilities must lie between 0 and 1) and with our information about the protocol of the experiments (the time duration $\Delta$ was estimated in our model and supposed to be lower than 240 seconds; the scaled size of the environment $\lambda$ was supposed between 1 and 20; handling times and vigilance bouts were measured in seconds). The best model was the 2D space with movement to the nearest with vigilance (see main text). 

To illustrate the importance of the fact that the parameters are estimated not only by fitting the expectation of the functional response but also its variance, we also fitted the models with and without vigilance using nonlinear model fitting  (``NonLinearModelFit'' in Mathematica 10.1 \citep{wolfram}), assuming a fixed variance independent of the parameters. Following this procedure, the model without vigilance had the lowest AIC and thus considered as the best model, in agreement with \cite{bakeretal2010}'s results and conclusions. However, this is due only to the fact that it has one less parameter to estimate, hence the AIC chooses the simplest model. 

As shown in the main text, the AIC of the model fitted with this latter procedure is lower than the model fitted with our framework (using Eq. \eqref{eq:LargeNumberLaw}). The estimation of the parameters are more constrained when both variance and expectation depends on the ecological parameters: the fitting procedure must find a parameters set satisfying the best trade-off between both the expectation and the variance of the functional response. When using the information from the variance of the data to estimate parameters, the different models show largely different log-likelihoods, showing that the difference of the AIC both procedures is not essentially due to the difference in the number of parameters to be estimated. Our results suggest a different conclusion than the one by \cite{bakeretal2010}: since the best model is with vigilance, we conclude that it significantly affects the feeding rate of grey partridges in the context of this experiment. Our results and conclusions illustrate the importance to use the information from the variance of the feeding rate, and not only its mean. Especially, our framework makes possible to express this variance as a function of the ecological parameters, because of the bottom-up approach: describing the behaviors and properties at the level of the individuals to make emerge functional responses at the macroscopic level.

\end{document}